\documentclass[aps,pre,twocolumn,superscriptaddress,amsfonts,10pt]{revtex4-1}

\usepackage{graphicx}
\usepackage{layouts}
\usepackage{pstool}
\usepackage{amssymb,amsmath}
\usepackage[hidelinks]{hyperref}
\usepackage[capitalise]{cleveref}
\EndPreamble
\usepackage[per-mode=symbol]{siunitx}
\usepackage[table]{xcolor}
\usepackage{subfig}

\newcommand{\mquad}{\quad\,\,}
\definecolor{RED}{RGB}{255,0,0}

\usepackage{cancel}

\definecolor{dark-green}{rgb}{0,0.5,0}
\newcommand{\DSA}[1]{#1}

\begin{document}

\title{Steady-state propagation speed of rupture fronts along one-dimensional frictional interfaces}

\author{David Sk{\aa}lid Amundsen}
\email{amundsen@astro.ex.ac.uk}
\affiliation{Astrophysics Group, School of Physics, University of Exeter, Exeter, EX4 4QL, United Kingdom}

\author{J{\o}rgen Kjoshagen Tr{\o}mborg}
\affiliation{Department of Physics, University of Oslo, Sem Sælands vei 24, NO-0316, Oslo, Norway}
\affiliation{Laboratoire de Tribologie et Dynamique des Syst\`emes, CNRS, Ecole Centrale de Lyon, 36 Avenue Guy de Collongue, 69134 Ecully, France}

\author{Kjetil Th{\o}gersen}
\affiliation{Department of Physics, University of Oslo, Sem Sælands vei 24, NO-0316, Oslo, Norway}

\author{Eytan Katzav}
\affiliation{Racah Institute of Physics, The Hebrew University, Jerusalem 91904, Israel}

\author{Anders Malthe-S{\o}renssen}
\affiliation{Department of Physics, University of Oslo, Sem Sælands vei 24, NO-0316, Oslo, Norway}

\author{Julien Scheibert}
\email{julien.scheibert@ec-lyon.fr}
\affiliation{Laboratoire de Tribologie et Dynamique des Syst\`emes, CNRS, Ecole Centrale de Lyon, 36 Avenue Guy de Collongue, 69134 Ecully, France}

\date{\today}


\begin{abstract}
The rupture of dry frictional interfaces occurs through the propagation of fronts breaking the contacts at the interface. Recent experiments have shown that the velocities of these rupture fronts range from quasi-static velocities proportional to the external loading rate to velocities larger than the shear wave speed. The way system parameters influence front speed is still poorly understood. Here we study steady-state rupture propagation in a one-dimensional (1D) spring-block model of an extended frictional interface, for various friction laws. With the classical Amontons--Coulomb friction law, we derive a closed-form expression for the steady-state rupture velocity as a function of the interfacial shear stress just prior to rupture. We then consider an additional shear stiffness of the interface and show that the softer the interface, the slower the rupture fronts. We provide an approximate closed form expression for this effect. We finally show that adding a bulk viscosity on the relative motion of blocks accelerates steady-state rupture fronts and we give an approximate expression for this effect. We demonstrate that the 1D results are qualitatively valid in 2D. Our results provide insights into the qualitative role of various key parameters of a frictional interface on its rupture dynamics. They will be useful to better understand the many systems in which spring-block models have proved adequate, from friction to granular matter and earthquake dynamics.
\end{abstract}


\maketitle

\section{Introduction} \label{sec:introduction}

Extended frictional interfaces under increasing shear stress eventually break and enter a macroscopic sliding regime. They do so through the propagation of a rupture-like micro-slip front across the whole interface. The propagation speed of such fronts is typically of the order of the sound speed in the contacting materials, which made them elusive to measurements until the rise of fast camera monitoring of frictional interfaces in the late 1990s. It is now well established that a whole continuum of front propagation speeds $v_\text{c}$ can be observed along macroscopic frictional interfaces, from intersonic (between the shear and compression wave speeds of the materials, $c_\text{s}$ and $c_\text{k}$, see e.g. \cite{Xia-Rosakis-Kanamori-Science-2004}) to quasi-static (proportional to the external loading rate, see e.g. \cite{Prevost2013probing, Romero2014probing}), going through sub-Rayleigh fronts ($v_\text{c} \lesssim c_\text{R}$ with $c_\text{R}$ the Rayleigh wave speed, see e.g. \cite{Rubinstein-Cohen-Fineberg-Nature-2004, BenDavid-Cohen-Fineberg-Science-2010, Schubnel2011photo-acoustic}) and slow but still dynamic fronts ($v_\text{c} \sim \text{\numrange{0.01}{0.1}} c_\text{R}$, see e.g. \cite{Rubinstein-Cohen-Fineberg-Nature-2004}). This huge variety in observed speed triggered the natural question of what the physical mechanisms underlying speed selection of micro-slip fronts are.

Experimentally, it has been shown that the larger the local shear to normal stress ratio $\tau / p$ just prior to rupture nucleation, the faster the local front speed \cite{BenDavid-Cohen-Fineberg-Science-2010}. This result is consistent with the observation that a larger shear stress promotes intersonic rather than sub-Rayleigh propagation \cite{Lu2007pulse-like,Nielsen-Taddeucci-Vinciguerra-GeophysJInt-2010}. The observed relationship between pre-stress and front speed has been reproduced in simulations for both fast \cite{Tromborg2011transition,Kammer2012propagation,Tromborg2014slow} and slow \cite{Tromborg2014slow} fronts. On other aspects of front speed, models are ahead of experiments, with various predictions still awaiting experimental verification. Among these are the following: (i) Models with simple Amontons-Coulomb (AC) friction \cite{Tromborg2011transition,Amundsen2012,Tromborg2014slow,Tromborg2015velocity} have shown that front propagation speed is controlled by $\bar{\tau}=\frac{\tau / p - \mu_\text{k}}{\mu_\text{s} - \mu_\text{k}}$, with $\mu_\text{s}$ and $\mu_\text{k}$ the local static and kinematic friction coefficients, $\bar{\tau}$ thus appearing as a generalisation of the parameter $\tau / p$ used to analyse the experimental data. (ii) A model with velocity-weakening AC friction \cite{Kammer2012propagation} has suggested that front speed is direction-dependent, with different speeds for fronts propagating with and against the shear loading direction. (iii) Two-dimensional (2D) spring-block models \cite{Tromborg2014slow,Tromborg2015velocity} and 1D continuous models \cite{Ohnaka1989cohesive,BarSinai2012slow} have shown that front speed $v_\text{front}$ is proportional to some relevant slip speed $v_\text{slip}$, with a relationship of the type $v_\text{front} \sim \frac{\text{shear modulus of the interface}}{\text{stress drop during rupture}} v_\text{slip}$.

Giving quantitative predictions of front speed is difficult for at least two reasons. First, any real interface is heterogeneous at the mesoscopic scales at which stresses can be defined (scale including enough micro-contacts), due both to intrinsic heterogeneities of the surfaces and to heterogeneous loading. Thus, even if the front speed was selected only locally, i.e. as a function of the local stresses and local static friction threshold, the front speed would still be varying with front position along the interface. Second, models actually show that front speed can have long transients \cite{Tromborg2015velocity} (extending over sizes comparable to that of the samples used in a number of experiments), even for carefully prepared homogeneous interfaces. This suggests that the instantaneous front speed does not only depend on local quantities, but rather on the slip dynamics along all the broken part of the interface and from all times after front nucleation. Here, we overcome these difficulties by (i) considering fronts propagating along homogeneous interfaces and (ii) focusing on front speed only in steady state propagation, i.e. after transients are finished. For the sake of simplicity and to enable analytical treatment, we use a 1D spring block model for the shear rupture of extended frictional interfaces, introduced in \cite{Amundsen2012} to study the propagation length of precursors to sliding \cite{Rubinstein-Cohen-Fineberg-PhysRevLett-2007,Maegawa-Suzuki-Nakano-TribolLett-2010,Scheibert-Dysthe-EPL-2010,Tromborg2011transition,Braun2014propagation,Kammer2014linear}. \DSA{Whereas the dynamics of multiple successive events in the macroscopic stick-slip regime of this model was discussed in detail in \cite{Amundsen2012}, here we focus on the steady state propagation of a single rupture front.} We show in the discussion that the main results obtained using the 1D model still hold in a 2D extension of the model.

We emphasise that fully dynamic (as opposed to cellular automata) spring-block (or spring-mass) models have previously been widely used in the literature to model not only friction \citep[see e.g.][]{Braun-Barel-Urbakh-PhysRevLett-2009,Maegawa-Suzuki-Nakano-TribolLett-2010,Tromborg2011transition, Amundsen2012,Capozza2012static,Tromborg2014slow} and earthquake dynamics \citep[see e.g.][]{Burridge-Knopoff-BSSA-1967,Carlson1994dynamics,Kawamura2012}, but also, among others, self-organized criticality in nonequilibrium systems with many degrees of freedom \citep[e.g.][]{deSousaVieira1992dynamic}, adsorbed chains at surfaces \citep[e.g.][]{Milchev1996static}, fluctuations in dissipative systems \cite[e.g.][]{Aumaitre2001power} or creep in granular materials \citep{Blanc2014on-and-off}.

\DSA{Rupture velocities in 1D spring-block models have been studied previously \cite{Langer1991,Myers-Langer-PRE-1993,Muratov-PRE-1999} in the framework of the Burridge-Knopoff (BK) model~\cite{Burridge-Knopoff-BSSA-1967}. In the BK model, a chain of blocks and springs is loaded uniformly from the top through an array of springs connected to a rigid rod. Note that this loading configuration differs from the one used in the present paper, in which the chain of blocks is loaded from one edge. In \cite{Langer1991}, the rupture speed of the BK model with velocity weakening friction was obtained in the case of a uniform loading exactly at the local slipping threshold. The model was found to have a range of possible propagation velocities among which one is selected dynamically. The rupture velocity was also found to be resolution-dependent. This resolution problem was solved in~\cite{Myers-Langer-PRE-1993} by introducing a short-wavelength cutoff, obtained by adding Kelvin viscosity to the model. Rupture velocities in the BK model with Amontons--Coulomb friction were studied in \citet{Muratov-PRE-1999}, and found to have a unique solution for any given value of the initial shear stress at the interface, with a well-defined continuum limit. We compare our results to those of \cite{Langer1991,Myers-Langer-PRE-1993,Muratov-PRE-1999} in \cref{sec:discussion}.}

The paper is organised as follows: We first describe our model and derive its non-dimensional form (\cref{sec:theModel}). We then present our results for the velocity of steady-state front propagation as a function of the pre-stress on the interface prior to rupture (\cref{sec:steady_state}), for three variants of the model. We start with a simple AC friction law and obtain a closed form equation for the front velocity. We then add either a bulk viscosity or an interfacial stiffness, and provide for each an approximate equation for front speed. In \cref{sec:discussion}, we discuss our results in the light of a 2D model. Conclusions are in \cref{sec:conclusions}. Four appendices provide additional mathematical details.

\section{Model description} \label{sec:theModel}

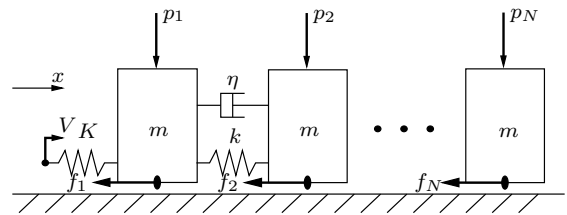
\begin{figure}[ht]
\centering
\begin{picture}(0,0)%
\includegraphics{{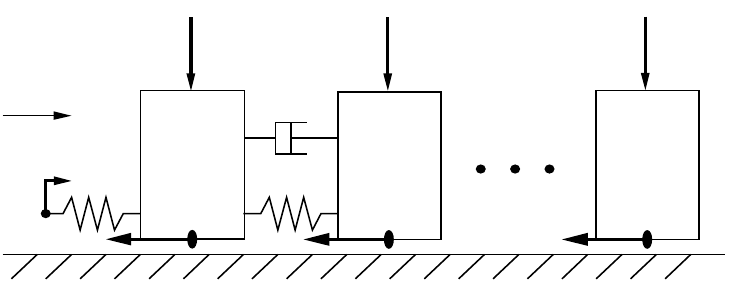}}%
\end{picture}%
\setlength{\unitlength}{3947sp}%
\begingroup\makeatletter\ifx\SetFigFont\undefined%
\gdef\SetFigFont#1#2#3#4#5{%
  \reset@font\fontsize{#1}{#2pt}%
  \fontfamily{#3}\fontseries{#4}\fontshape{#5}%
  \selectfont}%
\fi\endgroup%
\begin{picture}(3493,1347)(30,-73)
\put(380, 96){\makebox(0,0)[lb]{\smash{{\SetFigFont{8}{9.6}{\familydefault}{\mddefault}{\updefault}{\color[rgb]{0,0,0}$f_1$}%
}}}}
\put(1330, 96){\makebox(0,0)[lb]{\smash{{\SetFigFont{8}{9.6}{\familydefault}{\mddefault}{\updefault}{\color[rgb]{0,0,0}$f_2$}%
}}}}
\put(2569, 96){\makebox(0,0)[lb]{\smash{{\SetFigFont{8}{9.6}{\familydefault}{\mddefault}{\updefault}{\color[rgb]{0,0,0}$f_N$}%
}}}}
\put(455,409){\makebox(0,0)[lb]{\smash{{\SetFigFont{8}{9.6}{\familydefault}{\mddefault}{\updefault}{\color[rgb]{0,0,0}$K$}%
}}}}
\put(332,448){\makebox(0,0)[lb]{\smash{{\SetFigFont{8}{9.6}{\familydefault}{\mddefault}{\updefault}{\color[rgb]{0,0,0}$V$}%
}}}}
\put(290,761){\makebox(0,0)[lb]{\smash{{\SetFigFont{8}{9.6}{\familydefault}{\mddefault}{\updefault}{\color[rgb]{0,0,0}$x$}%
}}}}
\put(989,1159){\makebox(0,0)[lb]{\smash{{\SetFigFont{8}{9.6}{\familydefault}{\mddefault}{\updefault}{\color[rgb]{0,0,0}$p_1$}%
}}}}
\put(1935,1158){\makebox(0,0)[lb]{\smash{{\SetFigFont{8}{9.6}{\familydefault}{\mddefault}{\updefault}{\color[rgb]{0,0,0}$p_2$}%
}}}}
\put(3172,1163){\makebox(0,0)[lb]{\smash{{\SetFigFont{8}{9.6}{\familydefault}{\mddefault}{\updefault}{\color[rgb]{0,0,0}$p_N$}%
}}}}
\put(1405,379){\makebox(0,0)[lb]{\smash{{\SetFigFont{8}{9.6}{\familydefault}{\mddefault}{\updefault}{\color[rgb]{0,0,0}$k$}%
}}}}
\put(1389,741){\makebox(0,0)[lb]{\smash{{\SetFigFont{8}{9.6}{\rmdefault}{\mddefault}{\updefault}{\color[rgb]{0,0,0}$\eta$}%
}}}}
\put(1846,419){\makebox(0,0)[lb]{\smash{{\SetFigFont{8}{9.6}{\familydefault}{\mddefault}{\updefault}{\color[rgb]{0,0,0}$m$}%
}}}}
\put(901,419){\makebox(0,0)[lb]{\smash{{\SetFigFont{8}{9.6}{\familydefault}{\mddefault}{\updefault}{\color[rgb]{0,0,0}$m$}%
}}}}
\put(3098,416){\makebox(0,0)[lb]{\smash{{\SetFigFont{8}{9.6}{\familydefault}{\mddefault}{\updefault}{\color[rgb]{0,0,0}$m$}%
}}}}
\end{picture}%
\caption{Schematic of the model system. The slider is modelled as $N$ blocks with mass $m$ connected by springs of stiffness $k$. The trailing edge block (block $1$) is slowly driven through a loading spring of stiffness $K$. Each block $n$ is also submitted to a normal force $p_n$, a friction force $f_n$ and a viscous damping force $F_n^\eta$, which is described by the viscous coefficient $\eta$}.
\label{fig:modelSystem}
\end{figure}

We investigate the propagation of micro-slip fronts in the 1D spring-block model originally introduced by~\citet{Maegawa-Suzuki-Nakano-TribolLett-2010} to study the length of precursors to sliding. It has been later improved by us~\cite{Amundsen2012} to include a bulk viscosity and a friction law allowing for a finite stiffness of the interface. A schematic of this minimalistic model is given in Fig.~\ref{fig:modelSystem}. The slider is modelled as a chain of blocks with mass $m = M/N$ connected in series by springs with stiffness $k = (N-1)ES/L$, where $M$ is the total mass of the slider, $N$ is the number of blocks, $E$ is the Young's modulus, $S$ is the cross-section area and $L$ is the length of the slider. The applied normal force on each block $n$ is given by $p_n = F_\text{N}/N$, where $F_\text{N}$ is the total (uniformly) applied normal force. The tangential force $F_\text{T}$ is applied at the trailing edge of the system through a loading spring with stiffness $K$. One end of this spring is attached to the trailing edge block (block 1), while the other end moves at a (small) constant velocity $V$.

The equations of motion are given by
\begin{equation}
m \ddot{u}_n = F_n^k + F_n^\eta + f_n, \quad 1 < n \leq N,
\label{eq:eqOfMotion}
\end{equation}
where $u_n = u_n(t)$ is the position of block $n$ as a function of time relative to its equilibrium position (in the absence of any friction force) and $\ddot{}$ denotes the double derivative with respect to time $t$. The forces $F_n^k$, $F_n^\eta$ and $f_n$ are the total spring force, relative viscous force and the friction force on block $n$, respectively, and are given by
\begin{align}
F_n^k &= \left\{ \begin{array}{ll}
k(u_2 - u_1) + F_\text{T}, & n = 1 \\
k(u_{n+1} - 2u_n + u_{n-1}), & 2 \leq n \leq N-1 \\
k(u_{N-1} - u_N), & n = N,
\end{array} \right. \label{eq:F_n^k} \\
F_n^\eta &= \left\{ \begin{array}{ll}
\eta(\dot{u}_2 - \dot{u}_1), & n = 1 \\
\eta(\dot{u}_{n+1} - 2\dot{u}_n + \dot{u}_{n-1}), & 2 \leq n \leq N-1 \\
\eta(\dot{u}_{N-1} - \dot{u}_N), & n = N,
\end{array} \right. \label{eq:F_n^eta}
\end{align}
with the tangential load $F_\text{T}$ given by
\begin{equation}
F_\text{T} = K \left(Vt - u_1 \right) .
\label{eq:F_T(t)}
\end{equation}
In the following we will simply use the term ``viscosity'' to mean the viscous coefficient $\eta$. We consider two different functional forms for the friction force $f_n$, one corresponding to the rigid-plastic-like Amontons--Coulomb (AC) friction law, discussed in \cref{sec:AC}, and one to the elasto-plastic like friction law introduced in \citet{Amundsen2012} allowing for a finite stiffness of the interface, discussed in \cref{sec:interfacial_stiffness}. 


Before solving \cref{eq:eqOfMotion} it is instructive to rewrite it on a dimensionless form to derive the combination of parameters controlling the behaviour of the system. Here we derive the dimensionless equations of motion for a generic friction force $f_n$ and will later consider the two special cases discussed above, (i) AC friction in \cref{sec:AC} and (ii) with tangential stiffness of the interface in \cref{sec:interfacial_stiffness}.

We begin by eliminating the initial positions of all blocks, $u_n(0)$, from the block positions $u_n(t)$. Any movement can be described by $u_n'(t)$ defined by
\begin{subequations}
\begin{align}
u_n (t) &= u_n(0) + u_n'(t), \\
\dot{u}_n (t) &= \dot{u}_n'(t),\\
\ddot{u}_n (t) &= \ddot{u}_n'(t),
\label{eq:def:RelativePosition}
\end{align}
\end{subequations}
i.e. the position of a block is the position it had at $t = 0$ plus any additional movement $u_n'(t)$. The forces $F_n^k$, $F_n^\eta$ and $F_\text{T}$ then become
\begin{align}
F_n^k &= \left\{ \begin{array}{ll}
k(u_2' - u_1') + \tau_1 + F_\text{T}', & n = 1 \\
k(u_{n+1}' - 2u_n' + u_{n-1}') + \tau_n, & 2 \leq n \leq N-1 \\
k(u_{N-1}' - u_N') + \tau_N, & n = N,
\end{array} \right. \label{eq:F_n^k_prim} \\
F_n^\eta &= \left\{ \begin{array}{ll}
\eta(\dot{u}_2' - \dot{u}_1'), & n = 1 \\
\eta(\dot{u}_{n+1}' - 2\dot{u}_n' + \dot{u}_{n-1}'), & 2 \leq n \leq N-1 \\
\eta(\dot{u}_{N-1}' - \dot{u}_N'), & n = N,
\end{array} \right. \label{eq:F_n^eta_prim} \\
F_\text{T}' &= K(Vt - u_1'),
\label{eq:F_T_prim}
\end{align}
where we have introduced a new force $\tau_n$, the initial shear force, given by
\begin{equation}
\tau_n =  \left\{ \begin{array}{ll}
k(u_2(0) - u_1(0)) - Ku_1(0), & n = 1 \\
k(u_{n+1}(0) - 2u_n(0) + u_{n-1}(0)), & 2 \leq n \leq N-1 \\
k(u_{N-1}(0) - u_N(0)), & n = N .
\end{array} \right. \label{eq:tau_n}
\end{equation}

Next, we introduce our dimensionless variables, $\bar u_n = u_n'/U$ for block positions, $\bar t = t/T$ for time, and $\bar x = x/X$ for horizontal positions. Substituting these back into \cref{eq:eqOfMotion,eq:F_n^k_prim,eq:F_n^eta_prim,eq:F_T_prim}, yields our dimensionless equations of motion. We make the following choices for the scaling parameters $U$, $T$ and $X$:
\begin{align}
U &= \frac{(\mu_\text{s} - \mu_\text{k}) p_n}{k}, &
T &= \sqrt{m/k}, &
X &= a,
\label{eq:dim_scaling}
\end{align}
where $\mu_\text{s}$ and $ \mu_\text{k}$ are the static and kinetic friction coefficients in AC-like friction laws and $a=L/(N-1)$. The dimensionless equations of motion become
\begin{equation}
\ddot{\bar u}_n = \bar F_n + \bar F_n^{\bar \eta} + \bar f_n ,
\label{eq:EqOfMotionBar}
\end{equation}
with
\begin{align}
\bar F_n &= \left\{ \begin{array}{ll}
\bar u_2 - \bar u_1 + \bar F_\text{T}, & n = 1 \\
\bar u_{n+1} - 2\bar u_n + \bar u_{n-1}, & 2 \leq n \leq N-1 \\
\bar u_{N-1} - \bar u_N, & n = N,
\end{array} \right. \label{eq:FBar_n^k} \\
\bar F_n^{\bar \eta} &= \left\{ \begin{array}{ll}
\bar \eta(\dot{\bar u}_2 - \dot{\bar u}_1), & n = 1 \\
\bar \eta(\dot{\bar u}_{n+1} - 2\dot{\bar u}_n + \dot{\bar u}_{n-1}), & 2 \leq n \leq N-1 \\
\bar \eta(\dot{\bar u}_{N-1} - \dot{\bar u}_N), & n = N,
\end{array} \right. \label{eq:FBar_n^etaBar} \\
\bar f_n &= \frac{1}{kU} \left( f_n + \tau_n \right)
= \frac{\tau_n + f_n}{(\mu_\text{s} - \mu_\text{k}) p_n},
\label{eq:fBar} \\
\bar F_\text{T} &= \frac{T^2}{mU} K(Vt - u_1') = \bar K (\bar V \bar t - \bar u_1),
\label{eq:FBar_T}
\end{align}
where $\dot{}$ now denotes the derivative with respect to $\bar t$ and not $t$. Note that for convenience, and being of frictional origin, the initial shear force $\tau_n$ has been included in the effective dimensionless friction force $\bar f_n$ in \cref{eq:fBar} rather than in $\bar F_n$. The dimensionless relative viscosity is defined as
\begin{align}
\bar \eta &\equiv \frac{\eta}{\sqrt{km}} ,
\end{align}
and we have introduced
\begin{align}
\bar K &\equiv \frac{K}{k}, & \bar V &\equiv \frac{V}{U/T} .
\end{align}
The velocity of sound in this model is given by~\cite{Kittel-JWS-2005}
\begin{equation}
v_\text{s} = a \sqrt{\frac{k}{m}} ,
\label{eq:v_s}
\end{equation}
and in our dimensionless units this becomes
\begin{equation}
\bar v_s = a \sqrt{\frac{k}{m}} \frac{T}{X} = 1,
\end{equation}
which was the reason for our choice of $X$.

Looking at our new dimensionless set of equations it is clear that the number of parameters has been reduced. In addition to the dimensionless friction force $\bar f_n$, only $\bar K$, $\bar V$ and $\bar \eta$, the dimensionless driving spring constant, driving velocity and relative viscosity, respectively, will impact the evolution of the dimensionless block positions.

\section{Steady state front propagation} \label{sec:steady_state}

Most previous studies of rupture front propagation in spring-block models have initialised the models with the shear stresses set to zero. External loading is then applied, and the evolution of the systems in time is studied~\cite{Braun-Barel-Urbakh-PhysRevLett-2009,Maegawa-Suzuki-Nakano-TribolLett-2010,Scheibert-Dysthe-EPL-2010,Tromborg2011transition,Amundsen2012,Braun2014propagation}. The interface states at the time of nucleation of micro-slip fronts are selected naturally through the evolution of the system in time, mimicking the experimental setups of e.g.~\cite{Rubinstein-Cohen-Fineberg-PhysRevLett-2007,Maegawa-Suzuki-Nakano-TribolLett-2010}.

To facilitate systematic study of the front velocity in our system, as previously done in e.g.~\cite{Tromborg2014slow,Tromborg2015velocity}, we prepare a desired interface state at the time of rupture nucleation and look at the resulting rupture dynamics. The initial state is governed by the initial shear stresses, $\tau_n$, which is one of the important parameters in the effective friction force $\bar f_n$. In this paper we discuss steady-state front propagation, i.e. fronts propagating at a constant velocity $v_\text{c}$, and for this reason all surface properties, including $\tau_n$, are kept homogeneous throughout the interface. For convenience, the block index $n$ will therefore be dropped in the following discussion where possible. We also make the assumption that the driving spring constant $K$ is much smaller than the material spring constant $k$, i.e. $\bar K \ll 1$, which means $\bar F_\text{T}$ can be treated as constant during the front propagation. This assumption is valid both in models studied previously (e.g.~\cite{Amundsen2012,Tromborg2011transition,Maegawa-Suzuki-Nakano-TribolLett-2010}) and in experimental studies (e.g.~\cite{Maegawa-Suzuki-Nakano-TribolLett-2010,BenDavid-Cohen-Fineberg-Science-2010}).

Here we use both numerical and analytical tools to study steady-state rupture fronts in 1D spring-block models. In \cref{sec:AC} we measure the front speed in our model with AC friction. We compare the cases without and with bulk viscosity $\bar \eta$ and show that in a few special cases closed-form expressions for the front velocity as a function of model parameters may be obtained. In \cref{sec:interfacial_stiffness} we study the impact of a finite stiffness of the interface before concluding with some remarks on the complete model.

\subsection{Amontons--Coulomb friction} \label{sec:AC}

Perhaps the simplest dry friction law in wide-spread use is Amontons--Coulomb friction, which introduces static and dynamic friction coefficients $\mu_\text{s}$ and $\mu_\text{k}$, respectively. We impose this law locally on each block in our system as in \cite{Amundsen2012,Tromborg2011transition,Maegawa-Suzuki-Nakano-TribolLett-2010}, i.e. a block has to overcome a friction threshold $\mu_\text{s} p$ to start sliding, during which it experiences a force $\mu_\text{k} p$. The friction force $f_n$ is therefore given by
\begin{equation}
f_n = \left\{ \begin{array}{ll}
\leq \mu_\text{s} p, & \dot{u}_n = 0 \\
-\text{sgn}(\dot{u}_n) \mu_\text{k} p, & \dot{u}_n \neq 0 ,
\end{array} \right.
\label{eq:f_n_AC}
\end{equation}
where, when $\dot{u}_n = 0$, $f_n$ balances all other forces acting on block $n$. Blocks are assumed to repin to the track when their velocity becomes $0$ and will only start moving again if the sum of all forces, except the friction force, again reaches the static friction threshold $\mu_\text{s} p$.

For sliding blocks we insert \cref{eq:f_n_AC} into \cref{eq:fBar} and obtain the effective dimensionless friction force $\bar f_n$:
\begin{equation}
\bar f^\pm = \frac{\tau/p \mp \mu_\text{k}}{\mu_\text{s} - \mu_\text{k}} .
\label{eq:fBar_n_AC}
\end{equation}
Note the necessary separation into $\bar f^+$ and $\bar f^-$ at this point, where $\bar f^+$ applies if the block is moving in the positive direction and $\bar f^-$ if it is moving in the negative direction. This is related to the change in sign of the friction force $f_n$ as the block velocity changes between being positive and negative.

In the model, a front propagates in the following way: The driving force increases on block $1$ up to the local static friction level. As block $1$ moves, the tangential force on block $2$ increases, eventually reaching its static friction threshold, and starts to move. We interpret the successive onset of motion of blocks as the model equivalent of the micro-slip fronts observed in experiments and define the local front velocity as the distance between two blocks divided by the time interval between the rupture of two neighbouring blocks. We label these two blocks $n = i$ and $n = i+1$, and denote the time between the onset of motion of these two blocks by $\Delta t_i$. Since the material springs are very stiff, the distance between two neighbouring blocks can be approximated to be $a$, independent of time. We define the rupture velocity $v_\text{c}$ as
\begin{align}
v_\text{c} &= \frac{a}{\Delta t_i} , &
\bar v_\text{c} = v_\text{c} \frac{T}{X} &= \frac{v_\text{c}}{v_\text{s}},
\label{eq:v_c}
\end{align}
where we have used the velocity of sound in \cref{eq:v_s}.

A block begins to move if the forces on it reach the static friction threshold,
\begin{equation}
F_n^k + F_n^\eta = \pm \mu_\text{s} p,
\end{equation}
which in dimensionless units becomes
\begin{equation}
\bar F_n + \bar F_n^{\bar \eta} = \pm 1 - f^\pm .
\label{eq:block_move_crit}
\end{equation}

A rupture initiates when the total force on block $1$ exceeds the static friction threshold, while all other blocks are still unaffected. We initialise the system with $\tau + F_\text{T}' = \mu_\text{s} p$, i.e.
\begin{equation}
\bar \tau + \bar F_\text{T} = 1,
\label{eq:rupture_init_bar}
\end{equation}
where
\begin{equation}
\bar \tau \equiv \frac{\tau/p - \mu_\text{k}}{\mu_\text{s} - \mu_\text{k}} = \bar f^+.
\label{eq:tauBar}
\end{equation}
Note the definition of $\bar \tau$, which we will later show to be a very important model parameter. We exclusively consider positive initial shear forces, the maximum being restricted by the static friction threshold. Consequently, all values of $\bar \tau$, the dimensionless initial shear force, lie between $-\mu_\text{k}/(\mu_\text{s} - \mu_\text{k})$ and $1$.

To summarise, the equations of motion for the system are given by \cref{eq:EqOfMotionBar,eq:FBar_n^k,eq:FBar_n^etaBar,eq:fBar,eq:FBar_T,eq:fBar_n_AC}, and rupture initiates when \cref{eq:rupture_init_bar} is satisfied. We will now proceed by first considering the simplest case where $\bar \eta = 0$, before studying the effect of introducing a bulk viscosity.

\subsubsection{Model without bulk viscosity \texorpdfstring{($\bar \eta = 0$)}{(bar eta = 0)}} \label{sec:no_viscosity}

As in the model by~\citet{Maegawa-Suzuki-Nakano-TribolLett-2010}, we first take $\bar \eta = 0$ and apply AC friction locally at each block. Since, as discussed above, we keep $\bar F_\text{T}$ constant during rupture, and since its initial value is given by the rupture criterion (\cref{eq:rupture_init_bar}) only two parameters remain that control the front propagation, $\bar \tau = \bar f^+$ and $\bar f^-$. We first want to identify for which values of $\bar f^\pm$ steady-state ruptures can be supported.

\begin{figure}
\centering
\subfloat[Simulated front velocity as a function of position along the interface for several different values of $\bar \tau = \bar f^+$ obtained with $N = 200$. From bottom to top, $\bar \tau = -0.1, \, -0.01, \, 0.01, \, 0.1, \, 0.5$. Velocities are seen to approach a steady-state velocity for $\bar \tau > 0$, while for $\bar \tau < 0$ the fronts slow down and eventually arrest.\label{fig:steady_state_AC}]{\resizebox{\columnwidth}{!}{%
\includegraphics{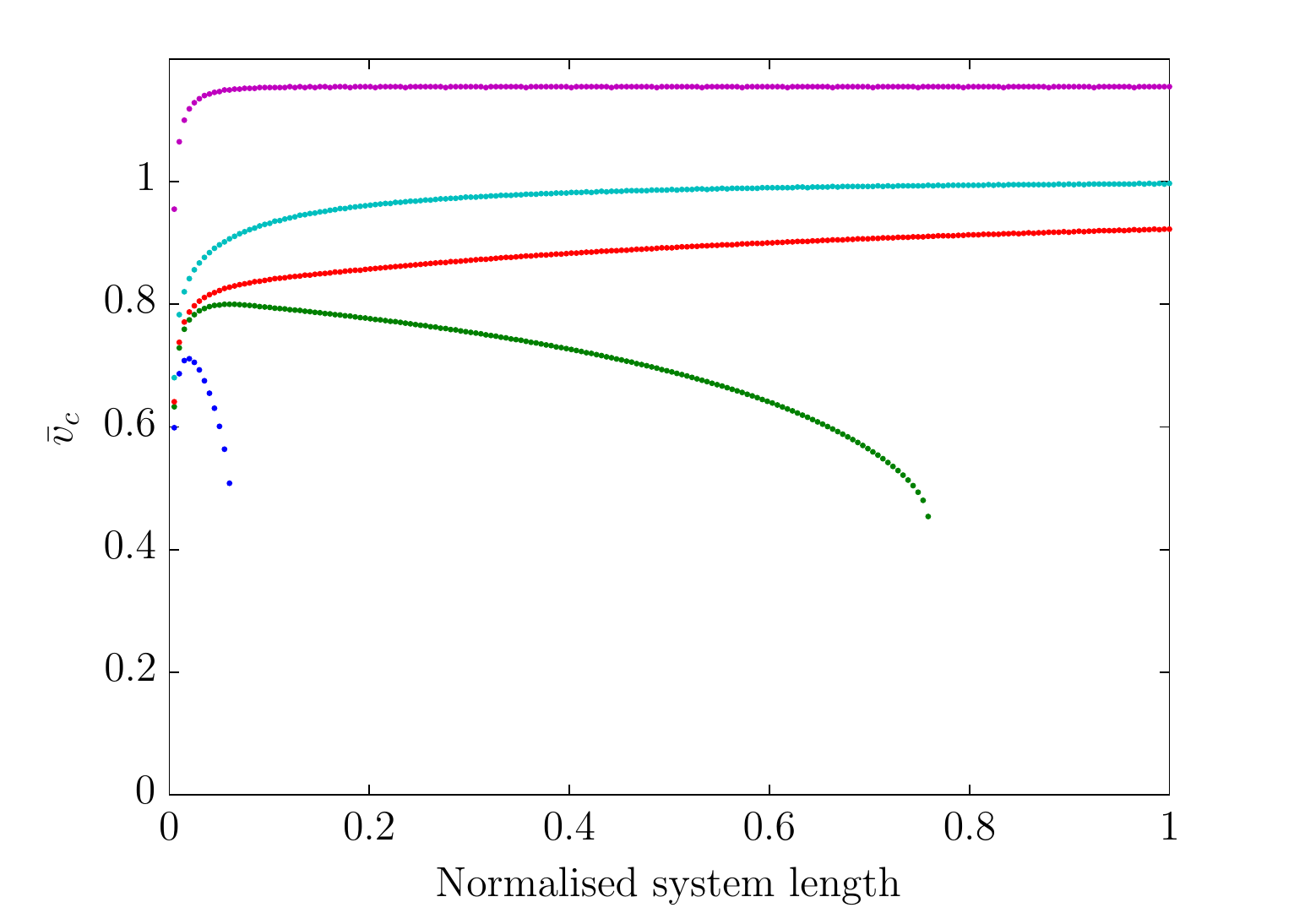}%
}}\\%
\subfloat[Transient length, defined as the block number where the front velocity reaches $\SI{97}{\percent}$ of the steady-state value, normalised by system size as a function of $\bar \tau$ obtained with $N = 1000$.\label{fig:transient_lengths}]{\resizebox{\columnwidth}{!}{%
\includegraphics{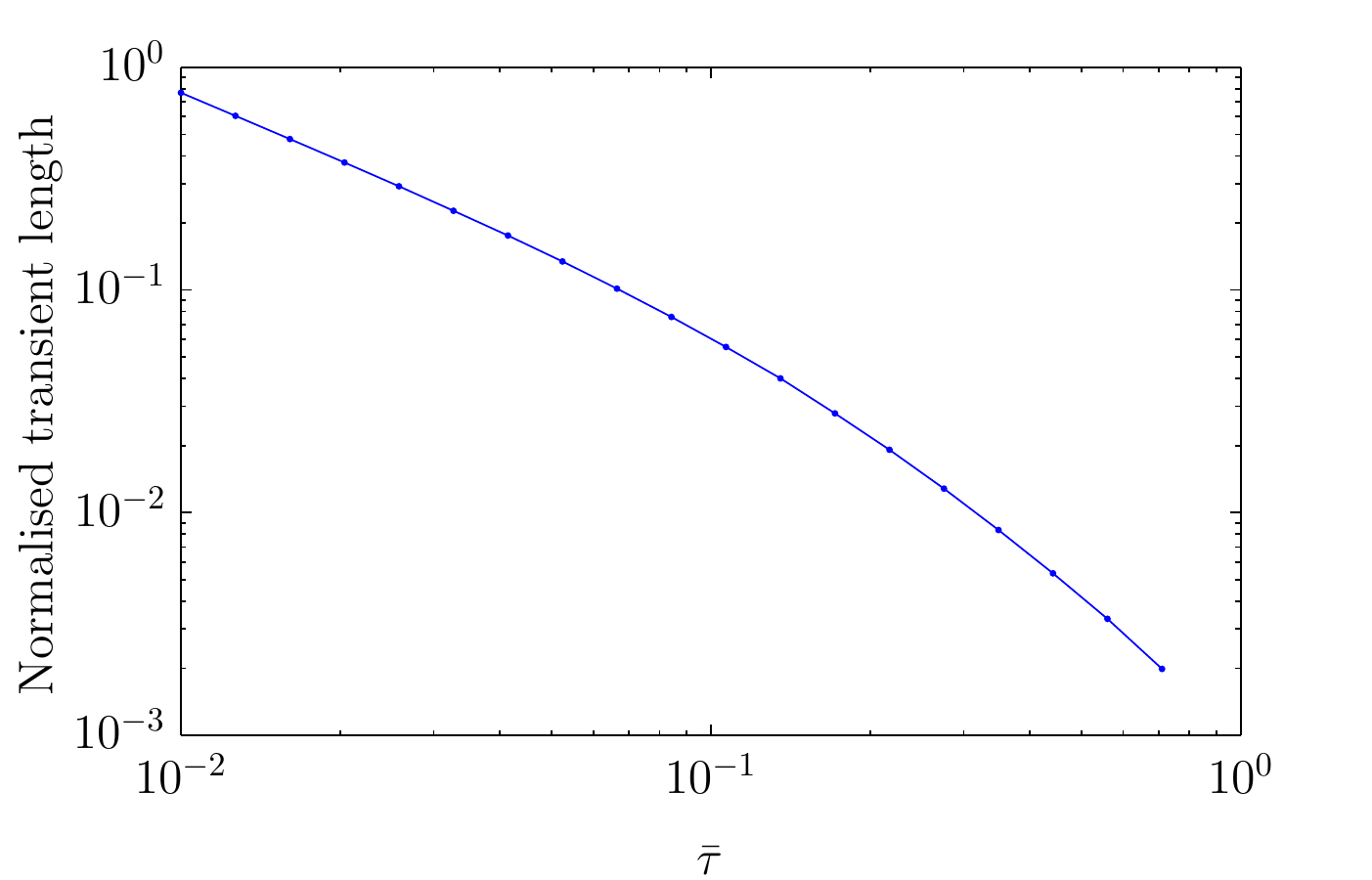}%
}}
\caption{Colour online. Rupture velocity as a function of position along the interface (a) and front velocity transient lengths (b) for various initial shear stress.}
\end{figure}

We have tried many different values for $\bar \tau$, some examples of which are shown in \cref{fig:steady_state_AC}. We observed that blocks never move in the negative direction for as long as the front propagates, which means that $\bar f^-$ becomes irrelevant, and we are left with only one parameter, $\bar \tau$, controlling front propagation. We have found that, for a steady-state rupture to be supported, $\bar \tau \geq 0$ is required. The natural restriction $\tau/p < \mu_\text{s}$ places another constraint on $\bar \tau$, and we can conclude that steady-state ruptures occur only if
\begin{equation}
0 \leq \bar \tau < 1.
\end{equation}

\begin{figure}
\centering
\resizebox{\columnwidth}{!}{%
\includegraphics{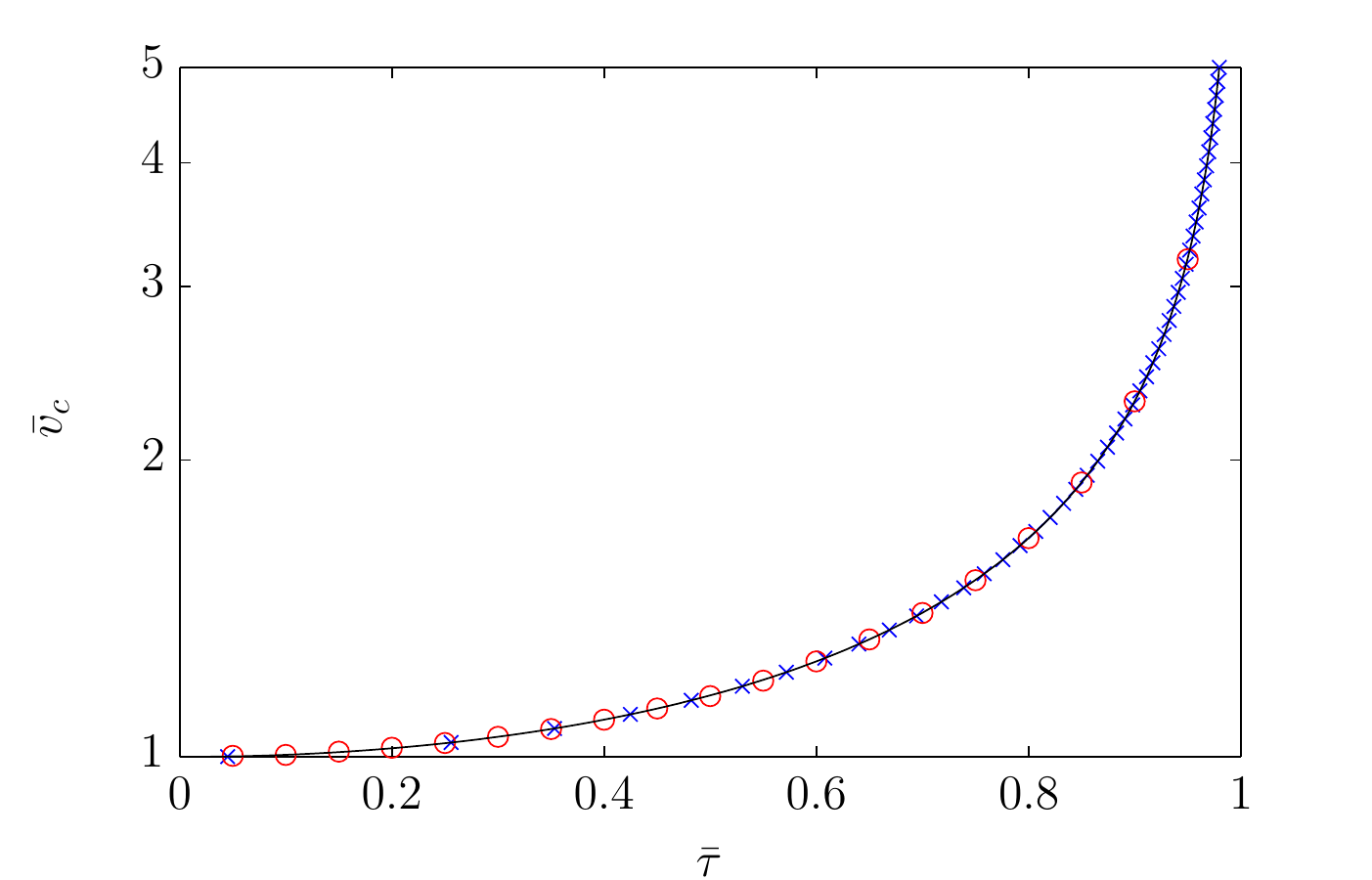}%
}
\caption{Colour online. Steady-state front velocity $\bar v_c$ as a function of initial stress, $\bar \tau$. $\bar v_c$ approaches $1$ as $\bar \tau \to 0$ and infinity as $\bar \tau \to 1$. Red circles are the extrapolated steady-state front velocities from simulations (as in \cref{fig:steady_state_AC}), blue crosses are the numerical solution of the steady-state equations (\cref{sec:steady_state_eqs_derive_AC}), and the solid black line is the analytical solution given by \cref{eq:tauBar_AC}.}
\label{fig:steady_state_AC_no_viscosity}
\end{figure}

The straightforward way to compute the steady-state rupture velocity as a function of $\bar \tau$ is to solve \cref{eq:EqOfMotionBar,eq:FBar_n^k,eq:FBar_n^etaBar,eq:fBar,eq:FBar_T,eq:fBar_n_AC} explicitly in time for a given value of $\bar \tau$. Fronts go through a transient, as seen in \cref{fig:steady_state_AC}, before reaching the steady-state velocity. As shown in \cref{fig:transient_lengths}, the transient length is strongly dependent on the pre-stress, $\bar \tau$, and extrapolation is necessary to estimate the final steady-state rupture velocity. We extrapolate the rupture velocity by fitting a first order polynomial to the curve $\bar v_c(1/n)$ for the last few ($\sim 50$) blocks towards the leading edge. These extrapolated steady-state velocities are plotted as a function of $\bar \tau$ in \cref{fig:steady_state_AC_no_viscosity} as red circles.

Alternatively, equations for the steady-state front velocity can be derived from \cref{eq:EqOfMotionBar,eq:FBar_n^k,eq:FBar_n^etaBar,eq:fBar,eq:FBar_T,eq:fBar_n_AC}. We provide this derivation in \cref{sec:steady_state_eqs_derive_AC}, with \cref{eq:eqOfMotionTransformed,eq:eqOfMotionTransformed_i,eq:initialCondition_i,eq:initialCondition} the final set of equations. The numerical scheme used to solve these equations is detailed in \cref{sec:steady_state_eqs_solve_num}. We show in \cref{fig:steady_state_AC_no_viscosity} the steady-state front velocity $\bar v_c$ as a function of $\bar \tau$ obtained by solving these equations numerically as blue crosses. This solution is seen to match the extrapolated front velocities obtained previously.

It is also possible to solve the equations for the steady-state rupture velocity, \cref{eq:eqOfMotionTransformed,eq:eqOfMotionTransformed_i,eq:initialCondition_i,eq:initialCondition}, analytically using an iterative approach. The solution technique is identical to the one used by \citet{Muratov-PRE-1999}, and we present the detailed calculation in \cref{sec:stead_state_eqs_solve_ana}. The result is a series expansion of $\bar \tau$ as a function of $z = 1/\bar v_c$, given in \cref{eq:tauBar_expansion_full}. In the present case with $\bar \eta = 0$, we get
\begin{equation}
\bar \tau = 1-\frac{z^2}{2}-\frac{z^4}{8}-\frac{z^6}{16}-\frac{5 z^8}{128}-\frac{7 z^{10}}{256} + \mathcal O\left(z^{12}\right),
\label{eq:tauBar_series_AC}
\end{equation}
which we recognise as the series expansion of
\begin{equation}
\sqrt{1 - z^2} = 1-\frac{z^2}{2}-\frac{z^4}{8}-\frac{z^6}{16}-\frac{5 z^8}{128}-\frac{7 z^{10}}{256}+ \mathcal O\left(z^{12}\right).
\end{equation}
Consequently, we have a closed-form expression for the front velocity as a function of the initial shear stress:
\begin{equation}
\bar \tau = \sqrt{1 - \bar v_c^{-2}} \quad \text{or} \quad \bar v_c = \frac{1}{\sqrt{1 - \bar \tau^2}}.
\label{eq:tauBar_AC}
\end{equation}
This solution is plotted as the black solid line in \cref{fig:steady_state_AC_no_viscosity}, which matches the numerical results perfectly.

To summarise, the steady-state front velocity $\bar v_c$ in the model with AC friction and $\bar \eta = 0$ only depends on the dimensionless initial shear stress, $\bar \tau$, and is given by \cref{eq:tauBar_AC}, plotted in \cref{fig:steady_state_AC_no_viscosity}. The front velocity increases with increasing $\bar \tau$, as expected. All steady-state front velocities are supersonic, with values approaching the sound velocity as $\bar \tau \to 0$ and infinity as $\bar \tau \to 1$. The latter is easily explained: as $\bar \tau \to 1$, every block will be infinitely close to the static friction threshold, and infinitesimally small movement of the neighbouring blocks is enough to set them into motion. The time between the triggering of neighbouring blocks will therefore approach $0$, causing the front velocity to approach infinity, see \cref{eq:v_c}. This is a known feature of spring-block models~\cite{Knopoff1997noncausality}. We discuss these results in more detail in \cref{sec:discussion}.

This model and the above results serve as our reference for investigating now the effect of a bulk viscosity and a tangential stiffness of the interface on the steady-state front velocity.

\subsubsection{The effect of a bulk viscosity} \label{sec:relative_viscosity}

In this section we study the effect of the relative viscosity $\bar \eta$, identical to the one used in \citet{Amundsen2012} to smooth grid-scale oscillations during front propagation~\cite{Myers-Langer-PRE-1993,Shaw-GRL-1994}. Physically it is a simple way of introducing energy dissipation that will occur during deformation. As in \cref{sec:no_viscosity} we have found that steady-state ruptures may occur if
\begin{equation}
0 \leq \bar \tau < 1,
\end{equation}
independent of the value of the viscosity, $\bar \eta$. Similarly we have also found that blocks exclusively move in the positive direction as the front propagates along the interface. Consequently, we have two parameters controlling front velocity in this system, $\bar \tau$ and $\bar \eta$.

\begin{figure}[tb]
\centering
\resizebox{\columnwidth}{!}{%
\includegraphics{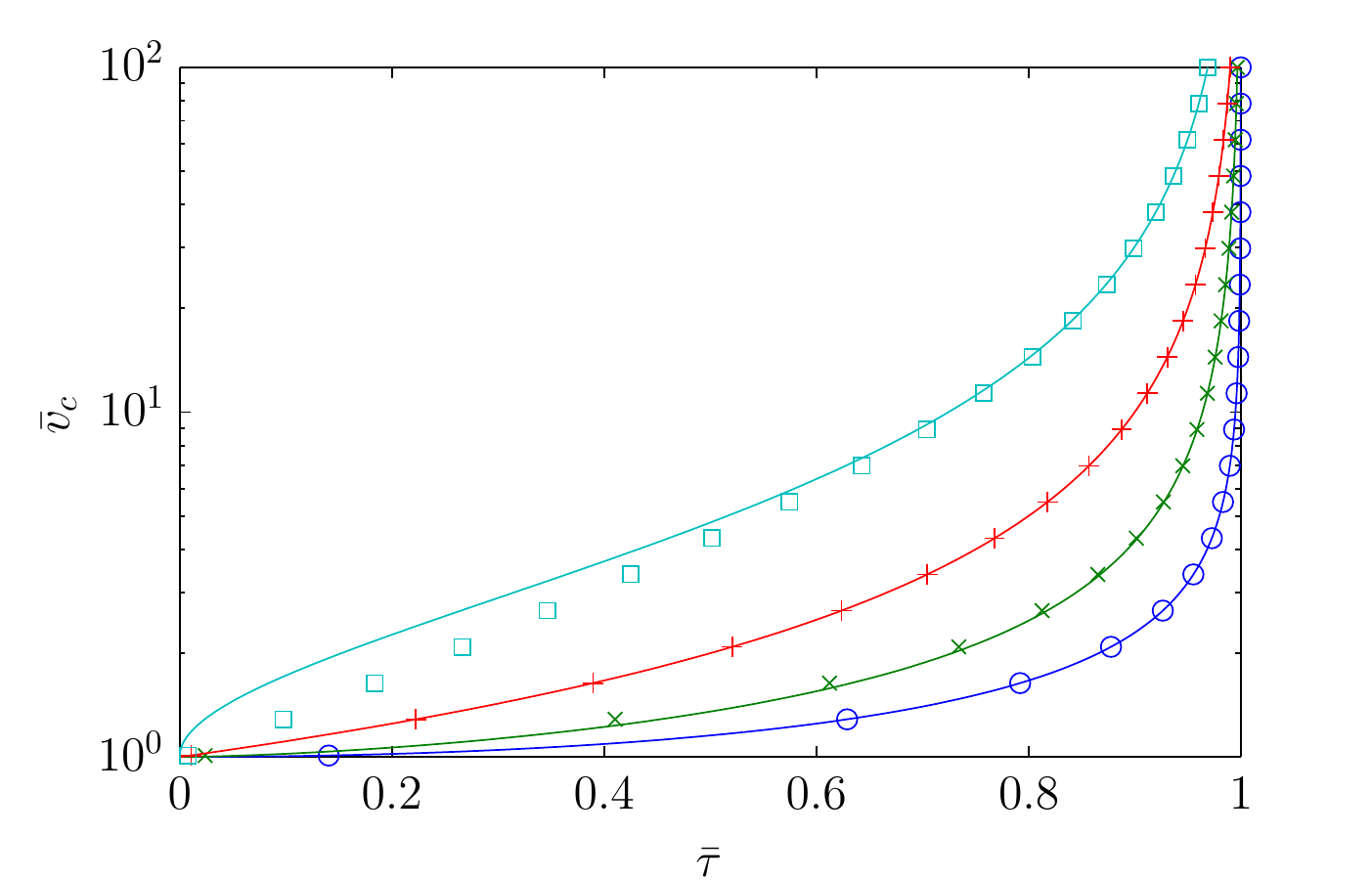}%
}
\caption{Colour online. Steady-state front velocity $\bar v_c$ as a function of initial shear stress, $\bar \tau$, for several different values of the relative viscosity $\bar \eta$. From bottom to top, $\bar \eta = 0$ (blue, circles), $\sqrt{0.1}$ (green, crosses), $1$ (red, pluses) and $\sqrt{10}$ (cyan, squares). The point markers are the numerical solutions of the steady-state equations, and the solid lines are the semi-empirical closed-form expression in \cref{eq:barTau_eta_empirical}.}
\label{fig:steady_state_AC_eta}
\end{figure}

The steady-state equations are solved numerically as in \cref{sec:no_viscosity} for several different values of $\bar \eta$, and we plot $\bar v_c(\bar \tau)$ in \cref{fig:steady_state_AC_eta}. As in the model with $\bar \eta = 0$, the steady-state equations can be solved analytically, and the result is a series expansion of $\bar \tau$ in terms of $z = 1/\bar v_c$ and $\bar \eta$. For brevity we do not reproduce this expansion here, it is given in \cref{eq:tauBar_expansion_full}.

Unfortunately we have been unable to find a closed-form expression for the front velocity as a function of $\bar \tau$ and $\bar \eta$ from \cref{eq:tauBar_expansion_full}. The special case $\bar \eta = 1$, however, yields
\begin{equation}
\bar \tau = 1 - z + \mathcal O\left(z^{12}\right),
\end{equation}
i.e.
\begin{equation}
\bar \tau = 1 - \frac{1}{\bar v_c} \quad \text{or} \quad \bar v_c = \frac{1}{1 - \bar \tau}.
\label{eq:tauBar_eta1}
\end{equation}
We use this to derive a semi-empirical expression for $\bar \tau(\bar v_c)$. From \cref{eq:tauBar_AC,eq:tauBar_eta1}, we can write
\begin{equation}
\bar \tau(\bar v_c, \bar \eta = 1) = \bar \tau(\bar v_c, \bar \eta = 0) \left( \frac{1 - \bar v_c^{-1}}{1 + \bar v_c^{-1}} \right)^{1/2} .
\end{equation}
The factor $\left[ (1 - z)/(1 + z) \right]^{1/2}$ can be thought of as a more general scaling factor including the $\bar \eta$ dependence, but taken at $\bar \eta = 1$. We have found that we can estimate the $\bar \eta$ dependence with an exponent of $\bar \eta/2$, which fits the numerical solution fairly well. We therefore propose the following approximation
\begin{align}
\bar \tau(\bar v_c, \bar \eta) &\approx \bar \tau(\bar v_c, \bar \eta = 0) \left( \frac{1 - \bar v_c^{-1}}{1 + \bar v_c^{-1}} \right)^{\bar \eta/2} \\
&= \sqrt{1 - \bar v_c^{-2}} \left( \frac{1 - \bar v_c^{-1}}{1 + \bar v_c^{-1}} \right)^{\bar \eta/2} .
\label{eq:barTau_eta_empirical}
\end{align}
We plot \cref{eq:barTau_eta_empirical} in \cref{fig:steady_state_AC_eta}, and it is seen to match the numerical solution perfectly for $\bar \eta = 0$ and $\bar \eta = 1$ as expected, and to yield acceptable accuracy for $0 \leq \bar \eta \leq 1$. The value of $\bar \eta = \sqrt{0.1}$ adopted by \citet{Amundsen2012} is within this range. The accuracy of the semi-empirical expression deteriorates for $\bar \eta > 1$, which corresponds to a regime where all waves are overdamped~\cite{Amundsen2012}.

In \cref{fig:steady_state_AC_eta} the relative viscosity is seen to not alter the limiting behaviour as $\bar \tau \to 0$ and $\bar \tau \to 1$, for which the front velocity still approaches $\bar v_c = 1$ and $\bar v_c \to \infty$, respectively. An overall increase of front velocities compared to the $\bar \eta = 0$ case, discussed in \cref{sec:no_viscosity}, is, however, seen.

The reason for this particular behaviour is that the relative viscosity serves to dampen out (reduce) relative motion between blocks. At the front tip, the rightmost moving block is increasing the load on the leftmost stuck block through the material spring connecting them and through the relative viscous force. This viscous force acts in the direction of movement of the moving block. This causes the static friction threshold to be reached sooner, and the stuck block starts moving earlier than it would have with a smaller value of $\bar \eta$. Note that the stuck block will act with an equal and opposite force on the moving block, slowing it down. This effect remains small, however, due to the large momentum of the blocks behind the front tip. Overall, $\Delta \bar t_n$, the time between the rupture of block $n$ and $n+1$, is reduced.

\subsection{Elasto-plastic like friction law} \label{sec:interfacial_stiffness}

As discussed in \citet{Amundsen2012}, in the model with AC friction, only the first block will experience the tangential loading force. This causes an unphysical resolution dependence in the model, which was improved considerably by introducing a finite tangential stiffness of the interface. In addition, the interface between the slider and the base is indeed elastic (see e.g.~\cite{Prevost2013probing}), a feature which is often accounted for in models using an ensemble of interface springs to model the micro-contacts binding the slider and base together~\cite{Braun-Peyrard-PhysRevLett-2008,Braun-Barel-Urbakh-PhysRevLett-2009,Tromborg2014slow,Thogersen2014history-dependent,Tromborg2015velocity}.

We introduce a tangential stiffness of the interface as in \citet{Amundsen2012} by modifying the static Amontons--Coulomb friction law to include springs between the blocks and the track. Each block bears one interface spring having a breaking strength equal to the static friction threshold $\mu_\text{s} p$. When a spring breaks, dynamic friction $\mu_\text{k} p$ applies until the spring reattaches when the block velocity becomes zero. The spring is reattached such that at the time of reattachment the total force on the block is zero.

In this section we study front velocity as a function of $\bar \tau$ and the stiffness of the interface springs, $k_\text{t}$. For attached blocks the friction force is given by
\begin{align}
f_n &= -k_\text{t} \left(u_n(t) - u_n^\text{stick}(t) \right) \notag \\
&= -k_\text{t} \left( u_n'(t) - {u_n^\text{stick}}'(t) \right) - k_\text{t} \left( u_n(0) - u_n^\text{stick}(0) \right),
\end{align}
where $u_n^\text{stick}(t)$ is the position of the attachment point of the spring. At $t = 0$ the total force on all blocks is zero, i.e. $f_n(t=0) = \tau_n$, and we get
\begin{equation}
f_n = -k_\text{t} \left( u_n'(t) - {u_n^\text{stick}}'(t) \right) - \tau_n.
\end{equation}
Using \cref{eq:fBar} we obtain an expression for the dimensionless friction force,
\begin{align}
\bar f_n &= \frac{\tau_n - k_\text{t} \left( u_n'(t) - {u_n^\text{stick}}'(t) \right) - \tau_n}{\mu_\text{s} p_n - \mu_\text{k} p_n} \\
&= -\frac{- k_\text{t} \left( u_n'(t) - {u_n^\text{stick}}'(t) \right)}{\mu_\text{s} p_n - \mu_\text{k} p_n} \\
&= - \bar k \left( \bar u_n(t) - \bar u_n^\text{stick}(t) \right), \label{eq:fBar_n_IS}
\end{align}
where $\bar k \equiv k_\text{t}/k$.

The rupture criterion is modified as it is a condition on the strength of the interface springs. It is given by
\begin{equation}
k_\text{t} \left( u_n(t) - u_n^\text{stick}(t) \right) = \mu_\text{s} p,
\end{equation}
which in dimensionless variables becomes
\begin{equation}
\bar k \left( \bar u_n(t) - \bar u_n^\text{stick}(t) \right) + \bar \tau_n = 1 .
\label{eq:InterfacialSpringRuptureCriterium}
\end{equation}
As discussed above, interface springs reconnect when the block velocity becomes zero and reconnect at zero total force:
\begin{align}
0 &= \bar F_n + \bar F_n^{\bar \eta} + \bar f_n \\
&= \bar F_n + \bar F_n^{\bar \eta} - \bar k \left( \bar u_n(t) - {\bar u}_n^\text{stick} (t) \right),
\end{align}
which yields
\begin{equation}
\bar u_n^\text{stick}(t^\text{stick}) = \bar u_n(t^\text{stick}) - \frac{\bar F_n(t^\text{stick}) + \bar F_n^{\bar \eta}(t^\text{stick})}{\bar k} ,
\end{equation}
where the only new parameter introduced is the dimensionless interface stiffness $\bar k = k_\text{t}/k$, and $t^\text{stick}$ is the time at which the block velocity becomes zero. $\bar u_n^\text{stick}(t)$ stays constant for $t > t^\text{stick}$ until the block reattaches after another detachment event.

For simplicity we investigate the behaviour of this model without the relative viscosity here, $\bar \eta = 0$, but this assumption is relaxed in \cref{sec:complete_model}. As in \cref{sec:no_viscosity} we have investigated when steady-state ruptures occur and found that it is again restricted to
\begin{equation}
0 \leq \bar \tau < 1,
\end{equation}
independent of the value of the interface stiffness, $\bar k$. Similarly we have also found that blocks exclusively move in the positive direction as the front propagates along the interface. Consequently, we have two parameters controlling the rupture velocity in this system, $\bar \tau$ and $\bar k$.

\begin{figure}
\centering
\resizebox{\columnwidth}{!}{%
\includegraphics{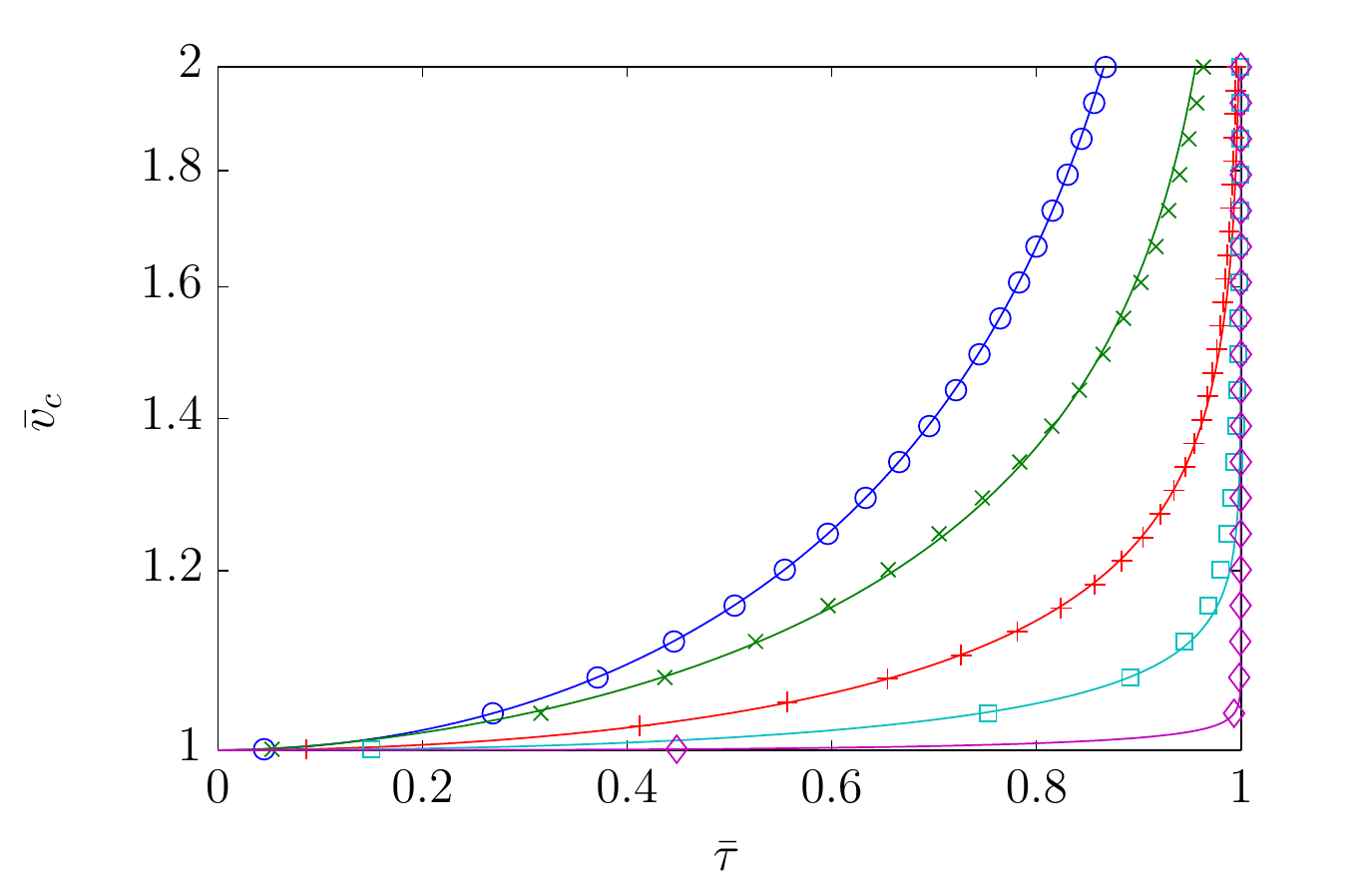}%
}
\caption{Colour online. Steady-state front velocity $\bar v_c$ as a function of initial shear stress, $\bar \tau$, for several different values of the interface stiffness $\bar k$. From top to bottom, $\bar k = \num{e3}, \, \num{e1}, \, \num{1}, \, \num{e-1}, \, \num{e-3}$. The crosses are the numerical solution, and the solid lines are the empirical closed-form expression in \cref{eq:barTau_kBar_empirical}.}
\label{fig:steady_state_IS}
\end{figure}

We solve the steady-state equations (derived in \cref{sec:steady_sate_eqs_derive_IS}) and show in \cref{fig:steady_state_IS} the front velocity as a function of the dimensionless initial shear force $\bar \tau$ for various values of the interface stiffness $\bar k$. In the limit $\bar k \to \infty$ the interface is infinitely stiff and the static friction law approaches AC's friction law. As $\bar k$ decreases the front velocity also decreases and $\bar v_c \to 1$ in the $\bar k \to 0$ limit where the interface is infinitely soft.

The rupture criterion, \cref{eq:InterfacialSpringRuptureCriterium}, is essentially a criterion for the displacement of a block relative to its attachment point. For a given pre-stress $\bar \tau$, as the interface stiffness $\bar k$ is reduced, blocks move a larger distance before detaching. In the limit $\bar k \to 0$ the rupture front will essentially become a displacement wave which moves with a velocity equal to the velocity of sound. This explains the behaviour of the model seen in the $\bar k \to 0$ limit.

Unfortunately we have not been able to obtain an analytical solution for the front velocity in the model with a tangential stiffness of the interface. Instead we have found the empirical expression
\begin{equation}
\bar \tau(\bar v_c, \bar k) = (1 - \bar v_c^{-n_1})^{1/n_2},
\label{eq:barTau_kBar_empirical}
\end{equation}
where the coefficients $n_1$ and $n_2$ are functions of $\bar k$, to yield satisfactory agreement with the numerical solutions. Best fit values of the coefficients $n_1$ and $n_2$, for the values of $\bar k$ in \cref{fig:steady_state_IS}, are given in \cref{tbl:kBar_exp}. These were obtained by solving the steady-state equations (\cref{sec:steady_sate_eqs_derive_IS}) numerically as described in \cref{sec:steady_state_eqs_solve_num} and fitting with \cref{eq:barTau_kBar_empirical} using a least squares method. The predictions made by \cref{eq:barTau_kBar_empirical} are shown as solid lines in \cref{fig:steady_state_IS}.

\begin{table}
\centering
\begin{tabular}{r|r|r}
$\bar k$ & $n_1$ & $n_2$ \\ \hline
$10^3$ & $2$ & $2$ \\
$10^1$ & $3.75$ & $1.7$ \\
$10^0$ & $7.86$ & $1.97$ \\
$10^{-1}$ & $20.8$ & $2.01$ \\
$10^{-3}$ & $106$ & $2.87$ \\
\end{tabular}
\caption{Best fit coefficients to be used in the empirical expression for the front velocity given in \cref{eq:barTau_kBar_empirical}. Note that the empirical approximation for $\bar k = \num{e3}$ is identical to \cref{eq:tauBar_AC}.}
\label{tbl:kBar_exp}
\end{table}

\subsection{Behaviour of the complete model} \label{sec:complete_model}

\begin{figure}
\centering
\resizebox{\columnwidth}{!}{%
\includegraphics{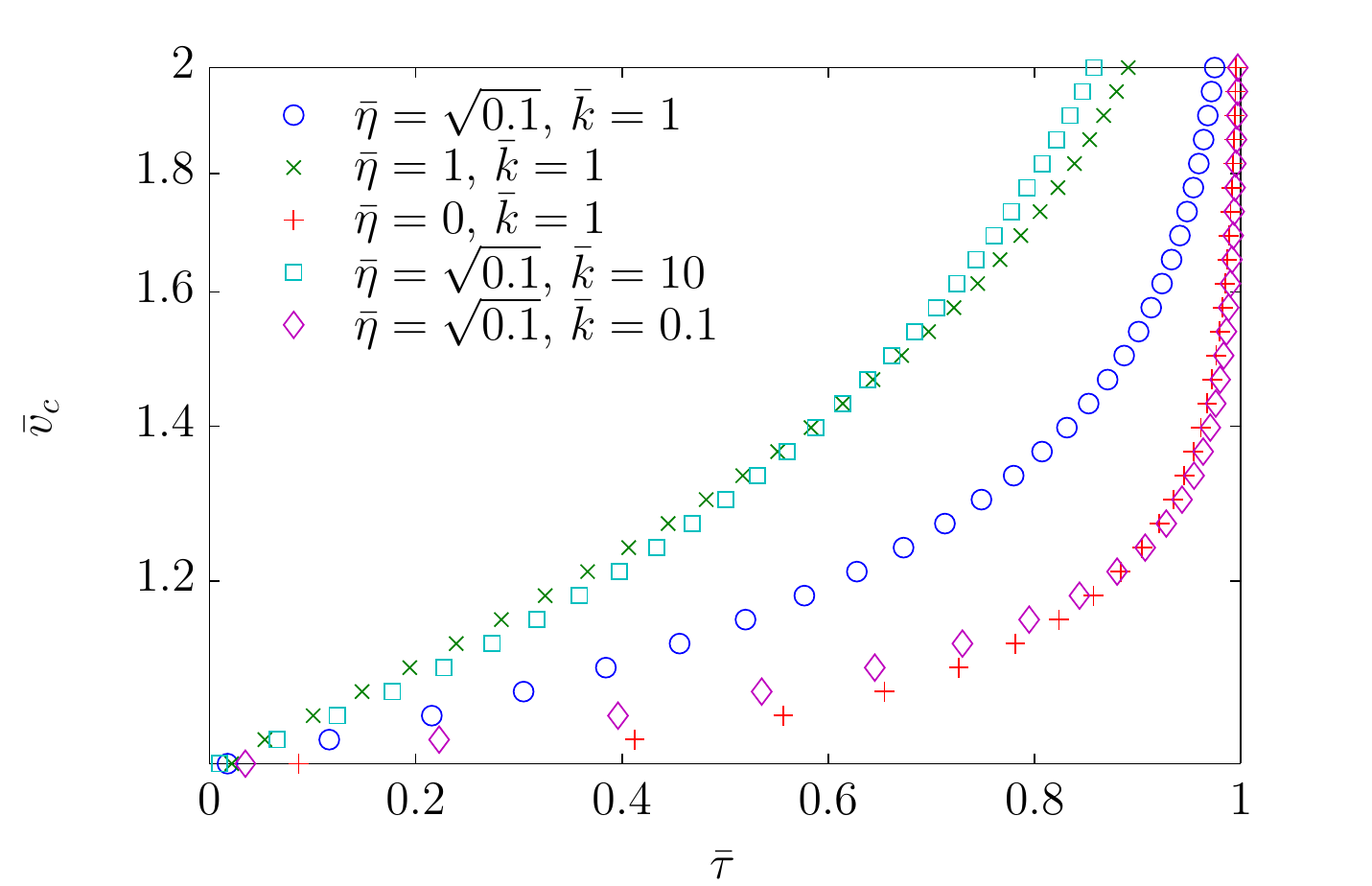}%
}
\caption{Colour online. Steady-state front velocity $\bar v_c$ as a function of initial shear stress, $\bar \tau$, for several different values of relative viscosity $\bar \eta$ and interface stiffness $\bar k$.} 
\label{fig:steady_state_complete} 
\end{figure}

\Cref{fig:steady_state_complete} shows the evolution of front velocity as a function of pre-stress in the complete model, i.e. where both a relative viscosity and a tangential stiffness of the interface are included. Several values of $\bar \eta$ and $\bar k$ are used, demonstrating that the complete model behaves in a way qualitatively consistent with the results of \cref{sec:relative_viscosity} and \cref{sec:interfacial_stiffness}. In particular, front speed increases both with increasing $\bar \eta$ and with increasing $\bar k$.

\section{Discussion} \label{sec:discussion}

We have found that, in general, the rupture front velocity increases with increasing pre-stress in our 1D spring-block-like models of extended frictional interfaces, as seen in \cref{fig:steady_state_AC_no_viscosity,fig:steady_state_AC_eta,fig:steady_state_IS,fig:steady_state_complete}. This is in agreement with observations on poly(methyl methacrylate) interfaces by \citet{BenDavid-Cohen-Fineberg-Science-2010}. We have also found the governing pre-stress parameter to be
\begin{equation}
\bar \tau \equiv \frac{\tau/p - \mu_\text{k}}{\mu_\text{s} - \mu_\text{k}}.
\end{equation}
That is, in addition to the pre-stress $\tau$ itself, the steady-state rupture velocity also depends on the local friction parameters $\mu_\text{s}$ and $\mu_\text{k}$, and on the applied normal load through $p$. The same parameter $\bar \tau$ has successfully been applied to 2D spring-block models to scale non-steady-state front velocities obtained with different model parameters (see Fig.~4b in \cite{Tromborg2011transition}, or \cite{Tromborg2015velocity}). It is also equivalent to the $S$ ratio used in the geophysical literature~\cite{Scholz-CUP-2002}, defined as
\begin{equation}
S = \frac{\sigma_\text{y} - \sigma_1}{\sigma_1 - \sigma_\text{f}} ,
\end{equation}
where $\sigma_\text{y}$, $\sigma_1$ and $\sigma_\text{f}$ are the yield stress, initial stress and sliding frictional stress, respectively. In terms of the parameters in the model discussed here, $\sigma_\text{y} = \mu_\text{s} p$, $\sigma_1 = \tau$, $\sigma_\text{f} = \mu_\text{k} p$, and we can express $S$ in terms of $\bar \tau$:
\begin{equation}
S = \frac{\mu_\text{s} - \tau/p}{\tau/p - \mu_\text{k}} = \frac{1}{\bar \tau} - 1 .
\end{equation}
The parameter $\bar \tau$ is therefore much more general than the derivation from the present model \DSA{alone} would indicate.


The second parameter of importance to steady-state ruptures is the relative viscosity parameter $\bar \eta$ discussed in \cref{sec:relative_viscosity}. It provides a simple way of introducing energy dissipation that will occur during deformation of the slider and also removes resolution-dependent oscillations in Burridge--Knopoff-like models~\citep{Amundsen2012,Knopoff-Ni-GJI-2001}, with the recommended value $\bar \eta = \sqrt{0.1}$. Note that the viscosity $\eta$ considered here is a bulk viscosity affecting the relative motion of blocks. It is thus qualitatively different from an interfacial viscosity that would affect the absolute motion of a single block on the track, as was sometimes introduced at the micro-junction level in multi-junction models (see e.g.~\cite{Braun-Barel-Urbakh-PhysRevLett-2009}) or directly at mesoscales as a velocity strengthening branch of the friction law at large slip velocities (see e.g.~\cite{BarSinai2015velocity-strengthening}). In our case, the friction force on a block in the sliding state is $\mu_\text{k}p$, independent of slip speed. At a given $\bar \tau$, increasing $\bar \eta$ increases the steady-state front velocity, see \cref{fig:steady_state_AC_eta}. As discussed in \cref{sec:relative_viscosity} this is due to the added shear force arising from the damping of relative motion between blocks. The particular choice $\bar \eta = \sqrt{0.1}$ (used in \cite{Amundsen2012,Tromborg2011transition}) is in \cref{fig:steady_state_AC_eta} seen to only modestly increase the front velocity compared to $\bar \eta = 0$.

The third and last parameter studied here is $\bar k = k_\text{t}/k$, the interface to bulk stiffness ratio, discussed in \cref{sec:interfacial_stiffness}. In \cref{fig:steady_state_IS} the limit $\bar k \to \infty$ is seen to yield Amontons--Coulomb-like behaviour, while decreasing $\bar k$ yields decreasing front velocities. In fact, as $\bar k \to 0$ the front velocity approaches the speed of sound as discussed earlier due to the front becoming a sound wave. These results should be relevant to various similar 1D and 2D models in which blocks are elastically connected to the base by springs~\cite{Braun-Barel-Urbakh-PhysRevLett-2009,Tromborg2011transition,Amundsen2012,Capozza2012static,Tromborg2014slow,Tromborg2015velocity}.

Recent simulations, in a 2D spring-block model with a friction law at the block scale emerging as the collective behaviour of many micro-junctions in parallel, have identified two different slip regimes for individual blocks~\cite{Tromborg2014slow}. A fast (inertial) slip regime is followed by a slow slip regime, controlled by the healing dynamics of the interface after rupture. Fronts driven by fast slip are fast inertial fronts, whereas fronts propagating when a significant portion of the slipping blocks are in the slow regime are slow~\cite{Tromborg2015velocity}. In this context, all fronts observed in the present 1D models are of the fast type.



Although as seen in \cref{fig:steady_state_AC} the transient front velocity is often sub-sonic, in our model, all steady-state fronts are supersonic, i.e. $\bar v_c > 1$. The fronts can propagate at arbitrarily large speeds as long as the pre-stress $\bar \tau$ is large enough. This has been discussed previously by \citet{Knopoff1997noncausality}. The velocity of sound in a 1D model is the longitudinal wave speed, while shear and Rayleigh waves do not exist. Nevertheless, we think it useful to point out that super-shear fronts have recently been observed in model experiments~\cite{BenDavid-Cohen-Fineberg-Science-2010}, and that in the geophysical community, such fronts have been both predicted theoretically~\cite{Freund1979,Day1982} and confirmed experimentally~\cite{Rosakis1999,Bouchon2003,Xia-Rosakis-Kanamori-Science-2004,Schubnel2011photo-acoustic}.

As seen in \cref{fig:transient_lengths}, and previously found in \cite{Tromborg2015velocity}, the initial transient in front speed before steady state is reached can be very long when $\bar\tau$ is close to zero. Because the dimensionless equations of motion do not change with the model resolution, it is clear that in this model, the length of the transients is given by a fixed number of blocks rather than a physical length, so we have overcome the problem of getting close to the steady state by performing simulations with a large number of blocks. In the other limit of $\bar\tau\to 1$, the transient length vanishes. We provide a demonstration of this result in \cref{sec:deriving_transient_length}.

To investigate the transient length's dependence on $\bar \tau$ for small values of $\bar \tau$ we initialise the system with a constant pre-stress as in \cref{fig:steady_state_AC_no_viscosity} and let the rupture propagate until its velocity has reached $\SI{97}{\percent}$ of the steady-state value. We define the point at which this happens as the transient length and plot in \cref{fig:transient_lengths} the transient length as a function of $\bar \tau$. For small values of $\bar \tau$ the transient length is very large.
%

The above considerations emphasise the fact that in spring-block models like the one studied here it is important to choose the resolution carefully: this choice will indeed select the physical length of transients in front dynamics. The size of each block should be equal to the screening length~\cite{Caroli-Nozieres-EurPhysJB-1998}, which in a purely elastic model is given by $\lambda_d \approx d^2/a$, where $d$ is the distance between micro-contacts and $a$ is the size of micro-contacts. For micrometer-ranged roughnesses, we expect $a\sim \SI{1}{\micro \metre}$ and $d \sim \text{\SIrange{10}{100}{\micro \metre}}$, yielding $\lambda \sim \text{\SIrange{0.1}{10}{\milli \metre}}$, i.e. in the millimeter range.
%
The typical horizontal length scale for extended lab-scale interfaces is $L \sim \SI{100}{\milli \metre}$, which yields $N \sim 100$, which is consistent with the number of blocks used here or in our previous studies~\cite{Tromborg2011transition,Amundsen2012,Tromborg2014slow,Tromborg2015velocity}.

\DSA{Let us now compare our results to those of previous studies of steady state rupture velocities. In the Amontons--Coulomb case, our solution $v_c(\bar \tau)$ takes the exact same form as the one found in~\cite{Muratov-PRE-1999} for the Burridge--Knopoff model with Amontons--Coulomb friction (compare \cref{eq:tauBar_expansion_full} and Eq.~(A14) in \cite{Muratov-PRE-1999}). As in~\cite{Muratov-PRE-1999}, we find a well-defined continuum limit where rupture velocities do not depend on the chosen resolution. Compared to the studies in~\cite{Langer1991,Myers-Langer-PRE-1993} of the Burridge--Knopoff model with velocity weakening friction, our results are qualitatively, although not quantitatively, similar. In particular, we also find that the rupture velocity increases with increasing shear prestress of the interface and with increasing values of the viscous coefficient. Note that \cite{Langer1991,Myers-Langer-PRE-1993,Muratov-PRE-1999} did not discuss the effect of an interfacial stiffness on rupture speed.}

\citet{Tromborg2014slow,Tromborg2015velocity} showed that, in their 2D model, the rupture and slip velocities are proportional. In 1D it is possible to derive a similar, exact relationship between the average slip speed and rupture velocity in the case of Amontons--Coulomb friction and no viscosity. This derivation is provided in \cref{sec:slip_vs_rupture}, where it is found that the local rupture velocity $\bar v_{c,i}$ at block $i$ in a simulation is related to the local average slip velocity $\dot{\bar u}_{i,\text{avg}}$ by
\begin{equation}
\bar v_{c,i} = \frac{\dot{\bar u}_{i,\text{avg}}}{1 - \bar \tau} .
\label{eq:slip_rescale}
\end{equation}
This relationship is demonstrated in \cref{fig:slip_vs_rupture}, where we have plotted the local rupture velocity as a function of the local average slip velocity for various values of $\bar \tau$ as measured in simulations similar to those seen in \cref{fig:steady_state_AC}. Rescaling the average slip velocity by $1/(1 - \bar \tau)$\DSA{, a straight line of unit slope is obtained.} For comparison with the rescaling formula found by \citet{Tromborg2014slow,Tromborg2015velocity}, it is instructive to write \cref{eq:slip_rescale} using dimensional quantities. \Cref{eq:v_c,eq:v_s,eq:dim_scaling} yield
\begin{equation}
v_{c,i} = \frac{\dot{u}_{i,\text{avg}} a k}{\mu_\text{s} p - \tau} = \frac{\dot{u}_{i,\text{avg}} ES}{\mu_\text{s}p - \tau},
\label{eq:slip_rescale_dim}
\end{equation}
where we have used $a = L/(N-1)$ and $k = (N-1)ES/L$ as in~\cite{Amundsen2012}, where $L$ is the length of the slider, $E$ is Young's modulus and $S$ is the \DSA{cross-sectional} area of the slider. This strongly resembles the rescaling formula in \cite{Tromborg2014slow,Tromborg2015velocity} which has the same denominator, while the characteristic force in the numerator is different due to the different interfacial laws applied.

An important question is whether the results obtained in the present 1D model can be extended to 2D models. To answer this question, we perform a series of simulations using the 2D model described in~\cite{Tromborg2014slow,Tromborg2015velocity}, with model parameters suitable for the study of steady state front propagation. In particular, the slider's length is twenty times larger than in~\cite{Tromborg2014slow}, so that front propagation has a chance to converge towards a steady state (the length of transients ranges from less than 40 blocks for $\bar\tau=0.95$ to longer than the system length for $\bar\tau=0.2$). We first choose a reference set of parameters, in which parameters are the same as in Table~S1 of \citep{Tromborg2014slow}, except $L=\SI{2.8}{\metre}$, $N_x=1140$, $M=\SI{1.5}{\kilo\gram}$, $F_N=\SI{3840}{\newton}$, $\bar\tau_\mathrm{trigger}= 0.95$, and $\alpha=\eta/160$. The rest of the settings \DSA{are} as follows: The width of the initial junction force distribution is zero, so the interface springs effectively act as a single spring per block. We checked that the mean junction slipping time $\bar t_R$, while important for slow fronts, does not affect these fast front results. Steady-state front velocity is estimated as in \cref{sec:no_viscosity}, by fitting a straight line to $\bar v_c(1/n)$ for the last ($\sim N_x/2$) blocks towards the leading edge (excluding the last $\sim 40$, which have $\bar v_c$ increasing due to an edge effect).

\begin{figure}
\centering
\resizebox{\columnwidth}{!}{%
\includegraphics{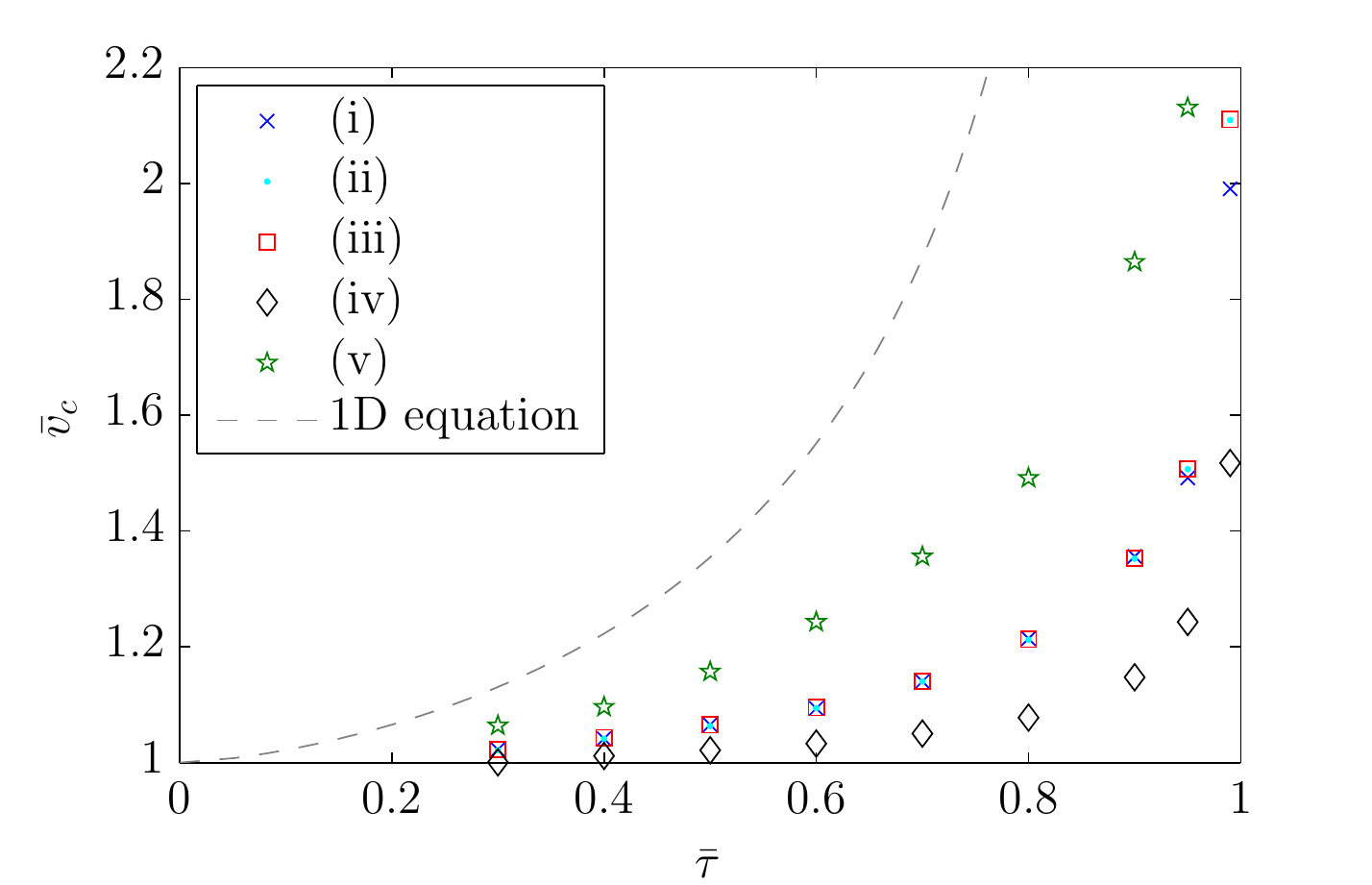}%
}
\caption{Colour online. Steady-state front velocity, $\bar v_c$, as a function of initial shear stress, $\bar\tau$, for simulations using the full 2D model of \citep{Tromborg2014slow}, for several different values of mass, damping coefficient, bulk stiffness and interface stiffness. Five sets of parameters were used: (i) Reference set; (ii) mass doubled and damping coefficient increased by a factor of $\sqrt{2}$, leaving $\bar\eta$ unchanged; (iii) bulk and interface stiffnesses halved, leaving $\bar k$ unchanged, (iv) damping coefficient and $\bar\eta$ reduced by a factor of $4$; and (v) interface stiffness and $\bar k$ increased by a factor of $10$. The dashed line plots \cref{eq:barTau_eta_empirical} for the reference value of $\bar\eta$.  Front speeds are normalized against the longitudinal bulk wave speed.}
\label{fig:2D}
\end{figure}

We first run simulations of the reference model for various values of $\bar\tau$ and plot the normalized steady state front speed $\bar v_c$ as a function of $\bar\tau$ in~\cref{fig:2D} (blue crosses). We observe that the results are qualitatively fully consistent with the behaviour in 1D, i.e. \cref{fig:steady_state_complete}. In particular, the steady state front speed is always supersonic, tends towards the longitudinal bulk wave speed for small prestress and \DSA{diverges} for large prestress.

We then show that \DSA{the two} control parameters identified in 1D, $\bar\eta$ and $\bar k$, are also controlling the 2D front speed. To do this, we \DSA{change} model parameters (slider's mass, damping coefficient, bulk and interfacial stiffnesses) in such a way that the rescaled parameters $\bar\eta$ and $\bar k$ are kept unchanged. \Cref{fig:2D} clearly shows (see cyan dots and red squares) that these changes do not affect the values of $\bar v_c$.

Finally, we show that changes to the values of either $\bar\eta$ or $\bar k$ induce variations in the $\bar v_c (\bar\tau)$ curve which are fully consistent with those observed in 1D (\cref{fig:steady_state_complete}): decreasing $\bar\eta$ decreases \DSA{the} front speed (see black \DSA{diamonds}) whereas increasing $\bar k$ increases \DSA{the front speed} (green stars). All these results indicate that \DSA{our} main \DSA{findings from} the 1D model are not specific to 1D, but also hold in 2D, with the same non-dimensional control parameters and all the same qualitative \DSA{characteristics}.

\section{Conclusions} \label{sec:conclusions}

We have systematically studied the quantitative dependence of the steady-state rupture front velocity on pre-stress, damping and stiffness of the interface in a 1D spring-block model. We find that front velocity changes significantly by changing any of these parameters, the result of which can be seen in \cref{fig:steady_state_AC_no_viscosity,fig:steady_state_AC_eta,fig:steady_state_IS,fig:steady_state_complete}.

Increasing pre-stress leads to increasing rupture velocities, in agreement with both experiments~\cite{BenDavid-Cohen-Fineberg-Science-2010} and 2D models (see discussion and ~\cite{Tromborg2011transition,Kammer2012propagation,Tromborg2015velocity}). Specifically, for the model with no viscosity and no finite stiffness of the interface, we derive a closed-form expression for the front velocity, given by \cref{eq:tauBar_AC}. The dimensionless pre-stress parameter found to be controlling front velocities, $\bar \tau = (\tau/p - \mu_\text{k})/(\mu_\text{s} - \mu_\text{k})$, also depends on the strength of the interface through the frictional parameters $\mu_\text{s}$ and $\mu_\text{k}$. It is essentially a version of the $S$ ratio often used by the geophysical community~\cite{Scholz-CUP-2002} and shown to also apply in 2D models~\cite{Tromborg2011transition,Tromborg2015velocity}.

Material damping affects the front velocity through the parameter $\bar \eta = \eta/\sqrt{km}$. Increasing values of $\bar \eta$ are seen to yield increasing front velocities caused by the additional shear force. A semi-empirical expression for the dependence of the front velocity on $\bar \eta$ and $\bar \tau$ is given in \cref{eq:barTau_eta_empirical}.

Front velocities are seen to decrease with decreasing tangential stiffness of the interface through the parameter $\bar k = k_\text{t}/k$, i.e. the ratio between the shear stiffness of the interface and the material stiffness of the slider. An empirical expression for this dependence is given in \cref{eq:barTau_kBar_empirical}. In fact, in the limit of a very soft interface compared to the material stiffness, steady-state rupture velocities are seen to approach the velocity of sound. The qualitative behaviour of all these parameters are seen to carry over to a model with both a finite stiffness of the interface and relative viscosity.

From \cref{fig:transient_lengths} it is clear that transients can become very long, especially for low pre-stresses. This, coupled with a heterogeneous interface where $\bar \tau$ can be negative, suggests that experimentally observed rupture fronts like those in e.g. \cite{Rubinstein-Cohen-Fineberg-Nature-2004,BenDavid-Cohen-Fineberg-Science-2010} may be dominated by transients. Direct comparison between these rupture fronts and the ones studied here may therefore not be possible, but the qualitative behaviour on parameters such as the internal damping and interface stiffness should remain valid.

\DSA{Also note that viscoelastic materials have been shown in finite element simulations to exhibit memory effects. Stress concentrations left at the arrest location of one precursory slip event are not erased by the following rupture~\cite{Radiguet2013}. This causes non-homogeneous initial stresses for subsequent events. An interesting direction for future work would be to investigate the transient speeds resulting from such complicated stress states.}

Despite the limitations of the model discussed above, it can provide valuable insight into rupture dynamics of frictional interfaces. We have shown here that the 1D results can be extended to a 2D model with the same interfacial law and bulk damping. Experimentally, it would be interesting to investigate the dependence of front velocities on the interface stiffness and bulk viscosity as they have been shown here to affect the rupture velocity significantly. Also, investigating the dependence of the front velocity on the system length would shed light on the influence of transients on observed ruptures.


\begin{acknowledgments}
This work was supported by the bilateral researcher exchange program Aurora (Hubert Curien Partnership), financed by the Norwegian Research Council and the French Ministry of Foreign Affairs and International Development (Grants No. 27436PM and No. 213213). K.T. acknowledges support from VISTA, a basic research program funded by Statoil, conducted in close collaboration with The Norwegian Academy of Science and Letters. J.S. acknowledges support from the People Programme (Marie Curie Actions) of the European Union's 7th Framework Programme (FP7/2007- 2013) under REA Grant Agreement 303871.
\end{acknowledgments}

\appendix

\section{Deriving the equations for the steady-state rupture velocity} \label{sec:steady_state_eqs_derive}

Here we derive the equations for steady-state rupture in the models discussed in this paper. We assume $\dot{u}_n (\bar t) > 0$ for all detached blocks, in agreement with the numerical solution of the equations for all cases where $\bar f^+ > 0$, as discussed in \cref{sec:steady_state}. The system is considered to be infinitely long and the tangential driving velocity much smaller than the front velocity, we can therefore ignore the system boundaries.

\subsection{Amontons--Coulomb friction} \label{sec:steady_state_eqs_derive_AC}

Here we derive the equations for the steady-state front velocity for the model with AC friction. As our starting point we use the dimensionless equations of motion, \cref{eq:EqOfMotionBar,eq:FBar_n^k,eq:FBar_n^etaBar,eq:fBar_n_AC,eq:rupture_init_bar}. Consequently, the controlling parameters in the equations of motion are $\bar \tau = \bar f^+$ and $\bar \eta$.

The equations of motion for moving blocks are given by
\begin{equation}
\begin{aligned}
\ddot{\bar u}_n &= \bar u_{n+1} - 2\bar u_n + \bar u_{n-1} \\
&\quad+ \bar \eta (\dot{\bar u}_{n+1} - 2\dot{\bar u}_n + \dot{\bar u}_{n-1}) + \bar \tau .
\end{aligned}
\label{eq:eqOfMotionDimless}
\end{equation}
To eliminate the dependence on $\bar \tau$ we introduce $\tilde u_n$ defined by
\begin{equation}
\bar u_n = \bar \tau \left( \tilde u_n + \bar t^2/2 \right),
\label{eq:uTilde_n}
\end{equation}
where the acceleration of blocks due to the force $\bar \tau$ is taken into account explicitly by the term $\bar \tau \bar t^2/2$ in \cref{eq:uTilde_n}. \Cref{eq:eqOfMotionDimless} simplifies to
\begin{equation}
\begin{aligned}
\ddot{\tilde u}_n &= \tilde u_{n+1} - 2\tilde u_n + \tilde u_{n-1} \\
&\quad+ \bar \eta (\dot{\tilde u}_{n+1} - 2\dot{\tilde u}_n + \dot{\tilde u}_{n-1}),
\end{aligned}
\label{eq:eqOfMotionTransformed}
\end{equation}
which is our final equation of motion for moving blocks.

If the front is propagating at a constant velocity,
\begin{equation}
\bar u_{n}(\bar t) = \bar u_{n+1} (\bar t + \Delta \bar t), \quad
\dot{\bar u}_{n}(\bar t) = \dot{\bar u}_{n+1} (\bar t + \Delta \bar t),
\label{eq:translationalInvariance}
\end{equation}
must hold, where $\Delta \bar t$ is the time between the triggering of two neighbouring blocks. It is therefore sufficient to consider the system in a time interval of length $\Delta \bar t$. We choose $\bar t \in \left[\bar t_i, \bar t_i + \Delta \bar t \right]$, where block $i$ begins to move at $\bar t = \bar t_i$. For convenience, and without loss of generality, we choose $\bar t_i = 0$. Using \cref{eq:uTilde_n} this yields
\begin{equation}
\tilde u_i (0) = 0, \quad
\dot{\tilde u}_i (0) = 0 ,
\label{eq:initialCondition_i}
\end{equation}
which is the initial condition for block $i$. For convenience we choose $i = 0$.

The initial condition for \cref{eq:eqOfMotionTransformed} can be obtained by evaluating \cref{eq:translationalInvariance} at $\bar t = 0$ and using \cref{eq:uTilde_n}. This yields
\begin{equation}
\begin{aligned}
\tilde u_{n}(0) &= \tilde u_{n+1} \left( \bar v_c^{-1} \right) + \frac{1}{2 \bar v_c^2}, \\
\dot{\tilde u}_{n}(0) &= \dot{\tilde u}_{n+1} \left( \bar v_c^{-1} \right) + \frac{1}{\bar v_c},
\end{aligned}
\label{eq:initialCondition}
\end{equation}
since
\begin{equation}
\Delta \bar t = \frac{a/X}{\bar v_c} = \bar v_c^{-1}
\label{eq:dt}
\end{equation}
from the definition of $X$, \cref{eq:dim_scaling}.

The equation of motion for all moving blocks is given by~\cref{eq:eqOfMotionTransformed}, but the equation of motion for block $i$ can be rewritten taking into account that block $i+1$ is stationary. Inserting $\bar u_{i+1} = 0$ into \cref{eq:uTilde_n} yields $\tilde u_{i+1} = -\bar t^2/2$, and using \cref{eq:eqOfMotionTransformed} with $n = i$ we have
\begin{equation}
\begin{aligned}
\ddot{\tilde u}_i &= \tilde u_{i-1} - 2\tilde u_i -\bar t^2/2 \\
&\quad+ \bar \eta (\dot{\tilde u}_{i-1} - 2\dot{\tilde u}_i - \bar t),
\end{aligned}
\label{eq:eqOfMotionTransformed_i}
\end{equation}

Solving \cref{eq:eqOfMotionTransformed,eq:eqOfMotionTransformed_i,eq:initialCondition_i,eq:initialCondition} result in $\tilde u_n(\bar t)$ for a given rupture velocity $\bar v_c$. This velocity is related to the parameter $\bar \tau$ through the rupture criterion, \cref{eq:block_move_crit}, at time $\bar t = 0$ for block $i = 0$. At $\bar t = 0$ we have $\bar u_0(0) = \bar u_1(0) = \dot{\bar u}_0(0) = \dot{\bar u}_1(0) = 0$, and using \cref{eq:block_move_crit,eq:FBar_n^k,eq:FBar_n^etaBar,eq:uTilde_n} in addition to $\bar \tau = \bar f^+$ we get
\begin{equation}
\tilde u_{-1}(0) + \bar \eta \dot{\tilde u}_{-1}(0) = \frac{1 - \bar \tau}{\bar \tau},
\end{equation}
or equivalently
\begin{equation}
\bar \tau = \frac{1}{\tilde u_{-1}(0) + \bar \eta \dot{\tilde u}_{-1}(0) + 1}
\label{eq:ruptureCriterion_AC}
\end{equation}
which relates the solution of \cref{eq:eqOfMotionTransformed,eq:eqOfMotionTransformed_i,eq:initialCondition_i,eq:initialCondition} for a given $\bar v_c$ to the corresponding $\bar \tau$. Note that these equations, in the case of $\bar \eta = 0$, reduce to the equations determining the front velocity in the Burridge--Knopoff model using AC friction and the approximations of slow and soft tangential loading as derived by \citet{Muratov-PRE-1999}.

\subsection{Including a finite tangential stiffness of the interface} \label{sec:steady_sate_eqs_derive_IS}

Here we derive the steady-state equations for the model including a tangential stiffness of the interface discussed in \cref{sec:interfacial_stiffness}. The equations of motion for sliding blocks are given by~\cref{eq:eqOfMotionTransformed} as their equation of motion is identical to the AC case. For stuck blocks we combine \cref{eq:EqOfMotionBar,eq:FBar_n^k,eq:FBar_n^etaBar,eq:fBar_n_IS,eq:uTilde_n} and get
\begin{equation}
\begin{aligned}
\ddot{\tilde u}_n &= \tilde u_{n+1} + \tilde u_{n-1} - 2\tilde u_n \\
&\quad+ \bar \eta (\dot{\tilde u}_{n+1}-2\dot{\tilde u}_{n} + \dot{\tilde u}_{n-1}) \\
&\quad- \bar k ( \tilde u_n + \bar t^2/2 ) - 1 ,
\end{aligned}
\label{eq:eqOfMotionAttached}
\end{equation}
where, without loss of generality, we have set $u_n^\text{stick} = 0$. For convenience we again let block $i = 0$ detach at $\bar t = 0$, and by the same argument as above these equations need only to be solved for $\bar t \in [0,\Delta \bar t]$. The initial conditions are the same as for the AC case, \cref{eq:initialCondition}, with the exception that only blocks far away from the rupture front keep a constant position. The initial conditions are therefore given by
\begin{equation}
\begin{aligned}
\tilde u_{n-1} (0) &= \tilde u_n (\bar v_c^{-1}) + \frac{1}{2 \bar v_c}, \\
\dot{\tilde u}_{n-1}(0) &= \dot{\tilde u}_n(\bar v_c^{-1}) + \frac{1}{\bar v_c}, \\
\tilde u_{n \to \infty} &= \dot{\tilde u}_{n \to \infty} = 0 .
\label{eq:initialCondition_IS}
\end{aligned}
\end{equation}
Using \cref{eq:InterfacialSpringRuptureCriterium,eq:uTilde_n} at $\bar t = 0$ with $\bar u_n^\text{stick} = 0$ we get 
\begin{equation}
\bar k \tilde u_n(0) = \frac{1 - \bar \tau}{\bar \tau} ,
\label{eq:ruptureCriterion_IS}
\end{equation}
which relates the solution of \cref{eq:eqOfMotionTransformed,eq:eqOfMotionAttached,eq:initialCondition_IS} for a given $\bar v_\text{c}$ to $\bar \tau$.

\section{Solving the equations for the steady-state rupture velocity} \label{sec:stead_state_eqs_solve}

Here we solve for the steady-state velocity analytically using the equation set derived in \cref{sec:steady_state_eqs_derive_AC} for the model with AC friction. In addition we describe a numerical solution procedure that we use to solve the steady-state equations for the models with either AC or elasto-plastic friction laws.

\subsection{Analytical solution of the Amontons--Coulomb steady-state rupture velocity equations} \label{sec:stead_state_eqs_solve_ana}

We solve \cref{eq:eqOfMotionTransformed,eq:eqOfMotionTransformed_i,eq:initialCondition_i,eq:initialCondition} using the iterative approach employed by \citet{Muratov-PRE-1999} where the solution is a power expansion in
\begin{equation}
z \equiv \bar v_c^{-1}.
\label{eq:z}
\end{equation}
If the front velocity is large, i.e. in the limit $z = \bar v_c^{-1} \to 0$, the distance required for block $0$ to move in order to initiate movement of block $1$ is negligible. Therefore, block $0$ is stationary in this limit. Also, when the rupture velocity is infinitely high, interactions between the blocks become less prominent because the time interval $\Delta \bar t$ becomes negligibly small.

To zeroth order in $z$, we therefore ignore the interaction terms in \cref{eq:eqOfMotionTransformed,eq:eqOfMotionTransformed_i}, which yields $\ddot{\tilde u}_n^{(0)} = 0$, i.e.
\begin{equation}
\tilde u_n^{(0)} = a_n^{(0)} \bar t + b_n^{(0)} ,
\end{equation}
where $a_n^{(0)}$ and $b_n^{(0)}$ are constants to be determined from the initial condition. Using \cref{eq:initialCondition} we get the two coupled difference equations
\begin{align}
a_n^{(0)} &= a_{n+1}^{(0)} + z, \\
b_n^{(0)} &= a_{n+1}^{(0)} z + b_{n+1}^{(0)} + \frac{1}{2} z^2,
\end{align}
where $a_0^{(0)} = b_0^{(0)} = 0$ from \cref{eq:initialCondition_i}. The solution is
\begin{align}
a_n^{(0)} &= -n z, \\
b_n^{(0)} &= \frac{n^2}{2} z^2 ,
\end{align}
and we therefore have
\begin{equation}
\tilde u_n^{(0)} = -n z \bar t + \frac{n^2}{2} z^2 , \quad
n \leq 0 .
\label{eq:u_n^0}
\end{equation}
This is the first iteration, giving the zeroth order solution to \cref{eq:eqOfMotionTransformed,eq:eqOfMotionTransformed_i,eq:initialCondition_i,eq:initialCondition}. The zeroth order relationship between $\bar v_c$ and $\bar \tau$ is obtained by using \cref{eq:u_n^0,eq:ruptureCriterion_AC}.

The first order solution is obtained by substituting $\tilde u_n$ and $\dot{\tilde u}_n$ using \cref{eq:u_n^0} into \cref{eq:eqOfMotionTransformed,eq:eqOfMotionTransformed_i}, integrating twice with respect to $\bar t$ and then using \cref{eq:initialCondition,eq:initialCondition_i} to eliminate the integration constants. This approach rapidly becomes cumbersome, so we have used \textsc{Mathematica} to go to higher orders. Here we simply give the solution:
\begin{widetext}
\begin{equation}
\begin{aligned}
\bar \tau &=
1 \,-\\
&\mquad \frac{z}{1!} \bar \eta \, + \\
&\mquad \frac{z^2}{2!} (\bar \eta - 1)(\bar \eta + 1) \, + \\
&\mquad \frac{z^3}{3!} \bar \eta (\bar \eta - 1)(\bar \eta + 1) \, - \\
&\mquad \frac{z^4}{4!} (\bar \eta - 1)(\bar \eta + 1) (7 \bar \eta^2 - 3) \, - \\
&\mquad \frac{z^5}{5!} \bar \eta (\bar \eta - 1)(\bar \eta + 1) (19 \bar \eta^2 - 11 ) \, + \\
&\mquad \frac{z^6}{6!} (\bar \eta - 1)(\bar \eta + 1) (229 \bar \eta^4-226 \bar \eta^2+45) \, + \\
&\mquad \frac{z^7}{7!} \bar \eta (\bar \eta - 1)(\bar \eta + 1) (995 \bar \eta^4-1154 \bar \eta^2+295) \, - \\
&\mquad \frac{z^8}{8!} (\bar \eta - 1)(\bar \eta + 1) (17151 \bar \eta^6-26837 \bar \eta^4+12349 \bar \eta^2-1575) \, - \\
&\mquad \frac{z^9}{9!} \bar \eta (\bar \eta - 1)(\bar \eta + 1) (102083 \bar \eta^6-178177 \bar \eta^4+95033 \bar \eta^2-14971) \, + \\
&\mquad \frac{z^{10}}{10!} (\bar \eta - 1) (\bar \eta + 1) (2294141 \bar \eta^8-4930384 \bar \eta^6+3640514 \bar \eta^4-1063816 \bar \eta^2+99225) \, + \\
&\mquad O\left( z^{11} \right).
\end{aligned}
\label{eq:tauBar_expansion_full}
\end{equation}
\end{widetext}

We conclude this section with a brief discussion of the validity of the solution in \cref{eq:tauBar_expansion_full}. It is valid only for $\bar \tau < 1$ due to the form of the rupture criterion. It is not valid for $\bar \tau = 0$ because of the transformation in \cref{eq:uTilde_n}. If $\bar \tau < 0$, then $\bar u_n$ becomes negative (combine \cref{eq:uTilde_n,eq:u_n^0}). As a result, the velocity must be negative, but it was assumed to be positive. \Cref{eq:tauBar_expansion_full} is consequently not valid for $\bar \tau \leq 0$, i.e., we get the constraints
\begin{equation}
0 < \bar \tau < 1,
\end{equation}
which is consistent with the numerical results in \cref{sec:steady_state}.

\subsection{Numerical solution procedure} \label{sec:steady_state_eqs_solve_num}

To solve the steady-state equations derived in \cref{sec:steady_state_eqs_derive} numerically, we used the following iterative scheme:
\begin{enumerate}
\item
Select the desired $z = \bar v_c^{-1}$ for which the value of $\bar \tau$ is to be found. An initial guess for the position and velocity of all blocks must be made at $\bar t = 0$, we used $\tilde u_n(0) = \dot{\tilde u}_n(0) = 0$ as the initial guess for all results presented here.
\item Solve
\Cref{eq:eqOfMotionTransformed,eq:eqOfMotionTransformed_i} or \cref{eq:eqOfMotionTransformed,eq:eqOfMotionAttached} using a numerical scheme for solving differential equations, e.g. the fourth order Runge--Kutta method.
\item
Calculate a new estimate for the initial conditions using \cref{eq:initialCondition_i,eq:initialCondition} or \cref{eq:initialCondition_IS} and the chosen front velocity. In the model with a tangential stiffness of the interface we have $\tilde u_{n \to \infty} = \dot{\tilde u}_{n \to \infty} = 0$, which is implemented as $\tilde u_{N} = \dot{\tilde u}_{N} = 0$ where $N$ is the number of blocks in the calculation. We find that $N = 100$ yields satisfactory results. Calculate $\bar \tau$ using \cref{eq:ruptureCriterion_AC} or \cref{eq:ruptureCriterion_IS}.
\item Repeat steps 2. and 3. until $\bar \tau$ has converged. The solution has converged when the difference in $\bar \tau$ between two iterations is less than some tolerance $\epsilon$. We use $\epsilon = \num{e-6}$.
\end{enumerate}
The functions $\bar v_c(\bar \tau)$ or equivalently $\bar \tau(\bar v_c)$ can be calculated by repeating the above steps for several values of $\bar v_c$.

\section{Deriving the \texorpdfstring{$\bar\tau \to 1$}{tauBar to 1} limit of transient length} \label{sec:deriving_transient_length}

Here we show that in the limit of $\bar\tau\to 1$ the transient length vanishes by considering the motion of the trailing edge block as it begins to move.

Using \cref{eq:EqOfMotionBar,eq:FBar_n^k,eq:FBar_n^etaBar,eq:fBar,eq:FBar_T} with AC friction [\cref{eq:tauBar}] the equation of motion for the trailing edge block becomes
\begin{equation}
\ddot{\bar u}_\text{TE} = - \bar u_\text{TE} + \bar \tau + \bar F_\text{T}
\end{equation}
where we have set $\bar \eta = 0$ for simplicity. As before we assume that $\bar F_\text{T}$ is independent of time. The initial condition for the trailing edge block is $\bar u_\text{TE}(\bar t = 0) = \dot{\bar u}_\text{TE}(\bar t = 0) = 0$ and rupture initiates when $\bar \tau + \bar F_\text{T} = 1$. This yields
\begin{equation}
\bar u_\text{TE}(\bar t) = 1 - \cos \bar t.
\end{equation}
The rupture criterion, \cref{eq:block_move_crit}, applied to the second block from the trailing edge yields
\begin{equation}
\bar u_\text{TE}(\bar t_1) = 1 - \cos \bar t_1 = 1 - \bar \tau,
\end{equation}
where $\bar t_1$ is the time at which the motion of the second block from the trailing edge is triggered. Noting that $z_1 = \bar t_1 = 1/\bar v_{c,1}$, we have
\begin{equation}
\bar \tau = \cos z = 1 - \frac{z^2}{2} + \frac{z^4}{24} + \mathcal O(z^6) .
\end{equation}
Comparing with \cref{eq:tauBar_series_AC}, the series of the steady-state rupture velocity, the first two terms are identical, i.e. as $\bar \tau \to 1$ ($z \to 0$), the rupture will instantly reach the steady-state velocity and the transient length approaches $0$. For smaller values of $\bar \tau$ (larger values of $z$), the transient length increases due to the deviations in the higher order terms.

\section{Slip speed vs rupture speed} \label{sec:slip_vs_rupture}

\begin{figure*}
\centering
\subfloat[Local rupture velocity $\bar v_c$ as a function of local slip velocity $\dot{\bar{u}}_\text{avg}$.]{\resizebox{\columnwidth}{!}{%
\includegraphics{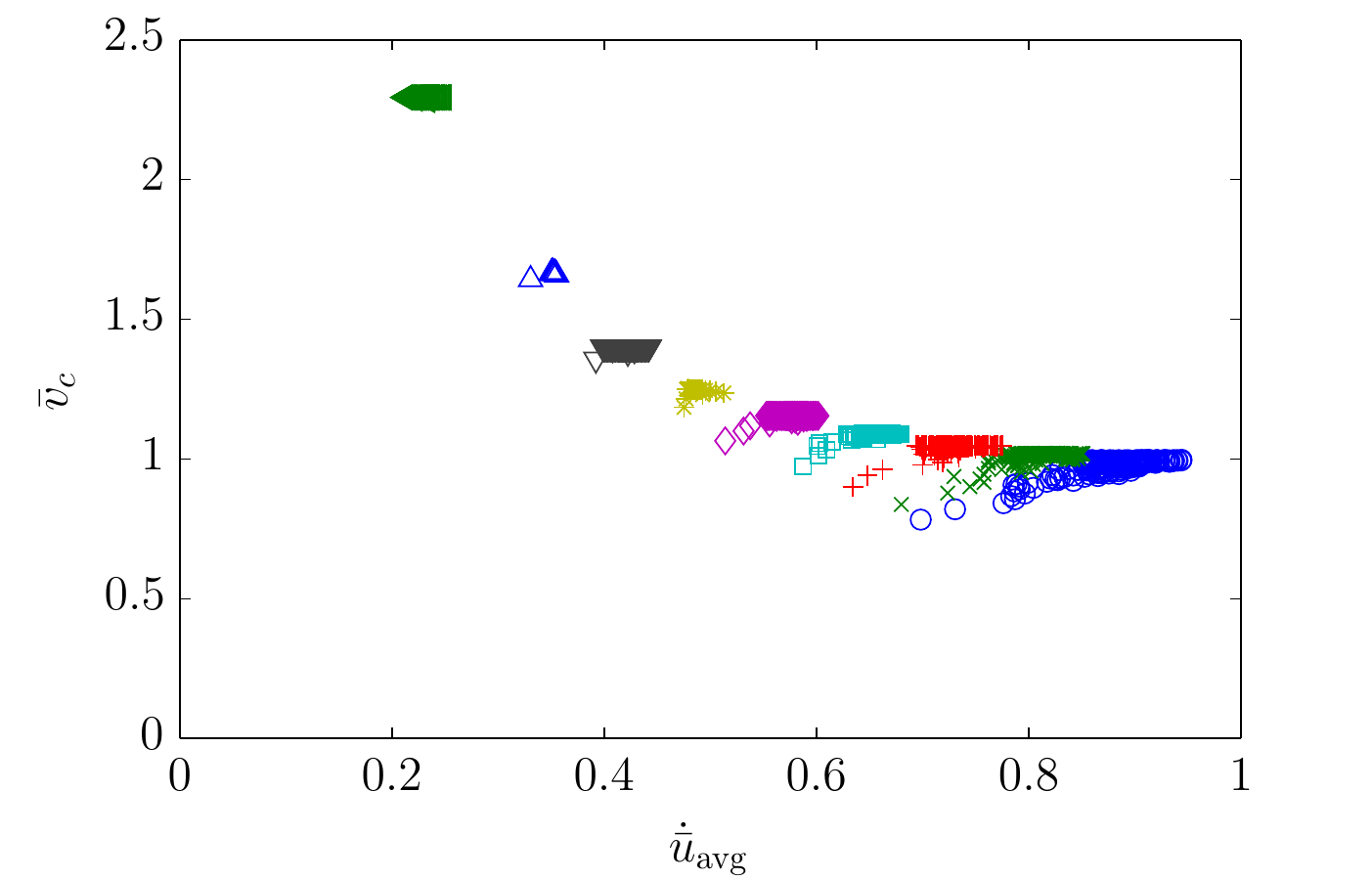}%
}}\quad
\subfloat[Local rupture velocity $\bar v_c$ as a function of the rescaled local slip velocity $\dot{\bar{u}}_\text{avg}/(1 - \bar \tau)$. The black line is a straight line through the origin with a slope of unity as predicted by \cref{eq:slip_rescale}.]{\resizebox{\columnwidth}{!}{%
\includegraphics{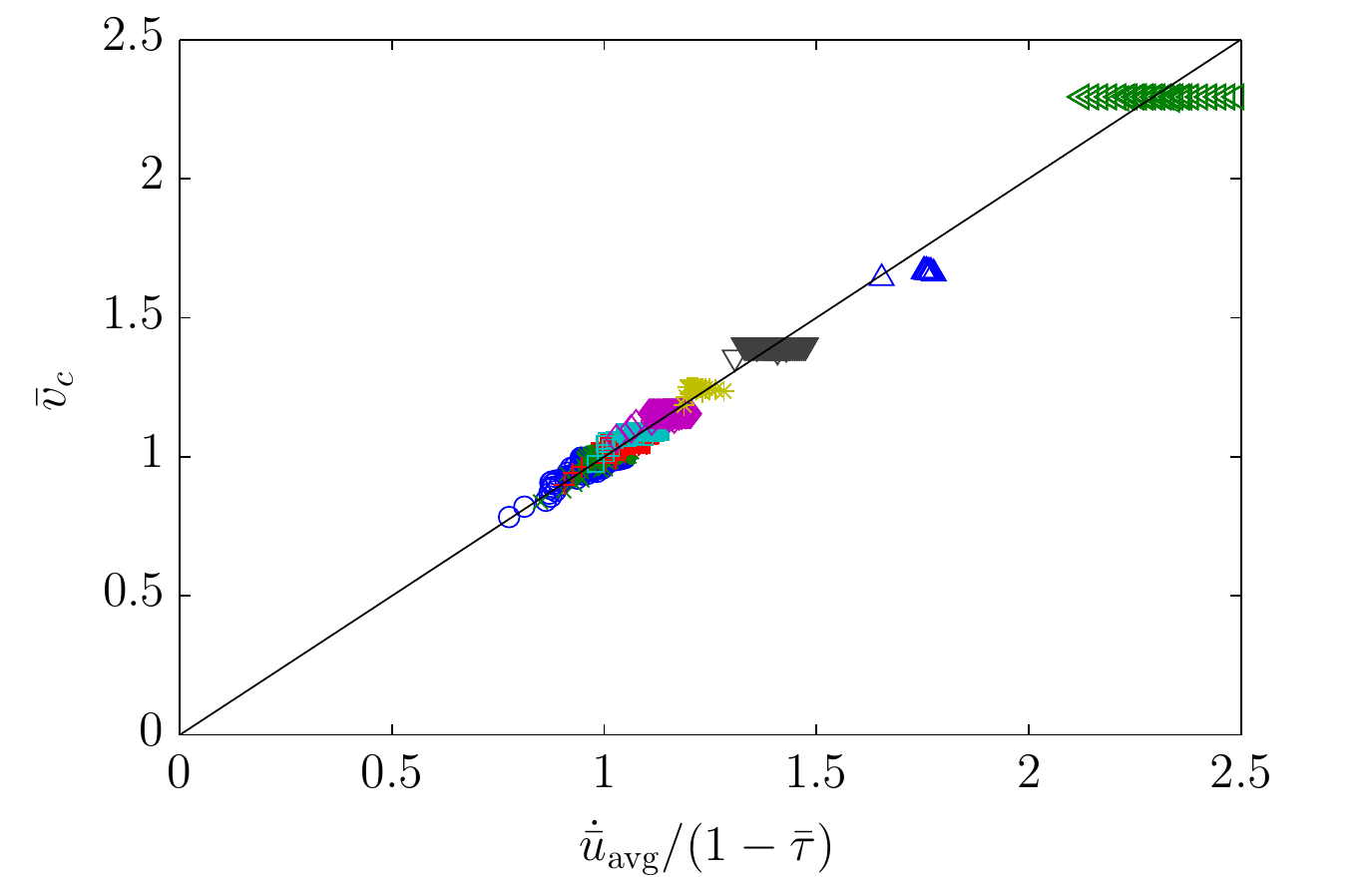}%
}}
\caption{Colour online. Rupture velocity and slip velocity are closely related. These results have been obtained using simulations as in \cref{fig:steady_state_AC} for $\bar \tau = 0.1,0.2, \dotsc, 0.9$.}
\label{fig:slip_vs_rupture}
\end{figure*}

Here we derive the relationship between the rupture and slip velocity in the model with Amontons--Coulomb friction and no viscosity. Consider block $i$, which started to move at time $\bar t = \bar t_i$, causing an increased shear force on block $i+1$. We only consider blocks moving in the positive direction, i.e. only $f_i^+ = \bar \tau$ is required. The local rupture velocity is given by \cref{eq:dt},
\begin{equation}
\bar v_{c,i} = \frac{1}{\Delta \bar t_i},
\end{equation}
where $\bar t_i = \bar t_{i+1} - \bar t_i$ is the time between the triggering of block $i$ and $i+1$. The position of block $i$ at $\bar t_i$ is $\bar u_i = 0$. The increase in shear force required for block $i+1$ to start moving is given by the rupture criterion, \cref{eq:block_move_crit}, and the position of block $i$ at $\bar t_{i+1}$ is therefore given by
\begin{equation}
\bar u_i(\bar t_{i+1}) = 1 - \bar \tau.
\end{equation}
The average slip velocity is consequently
\begin{equation}
\dot{\bar u}_{i,\text{avg}} = \frac{\bar u_i(\bar t_{i+1}) - \bar u_i(\bar t_i)}{\Delta \bar t_i}
= \frac{1 - \bar \tau}{\Delta \bar t_i},
\end{equation}
and the rupture and slip velocities are related by
\begin{equation}
\bar v_{c,i} = \frac{\dot{\bar u}_{i,\text{avg}}}{1 - \bar \tau} .
\tag{\ref{eq:slip_rescale}}
\end{equation}
This relationship is demonstrated in \cref{fig:slip_vs_rupture}, where we have plotted the local rupture velocity as a function of the local average slip velocity for various values of $\bar \tau$ as measured in simulations (similar to the simulations shown in \cref{fig:steady_state_AC}). Rescaling the average slip velocity by $1/(1 - \bar \tau)$, a straight line with unit slope is obtained.

\bibliography{bibPRE}

\end{document}